\pdfoutput=1

\newcommand\apjcls{1}
\newcommand\aastexcls{2}
\newcommand\othercls{3}

\newcommand\papercls{\aastexcls}
\documentclass[tighten, times, preprint2]{aastex6}  

\if\papercls \apjcls
\usepackage{apjfonts}
\else\if\papercls \othercls
\usepackage{epsfig}
\fi\fi
\usepackage{ifthen}
\usepackage{natbib}
\usepackage{amssymb, amsmath}
\usepackage{mathrsfs}
\usepackage{bm}
\usepackage{appendix}
\usepackage{etoolbox}
\usepackage[T1]{fontenc}
\usepackage{paralist}

\if\papercls \apjcls
\newcommand\aas{\ref@jnl{AAS Meeting Abstracts}}
\newcommand\dps{\ref@jnl{AAS/DPS Meeting Abstracts}}
\newcommand\maps{\ref@jnl{MAPS}}
\else\if\papercls \othercls
\usepackage{astjnlabbrev-jh}
\fi\fi

\if\papercls \aastexcls
\hypersetup{citecolor=blue, 
            linkcolor=blue, 
            menucolor=blue, 
            urlcolor=blue}  
\else
\usepackage[
bookmarks=true,           
bookmarksnumbered=true,   
colorlinks=true,          
citecolor=blue,           
linkcolor=blue,           
menucolor=blue,           
urlcolor=blue,            
linkbordercolor={0 0 1},  
pdfborder={0 0 1},
frenchlinks=true]{hyperref}
\fi

\if\papercls \othercls

\else

\fi

\providecommand{\adsurl}[1]{\href{#1}{ADS}}

\makeatletter
\patchcmd{\NAT@citex}
  {\@citea\NAT@hyper@{%
     \NAT@nmfmt{\NAT@nm}%
     \hyper@natlinkbreak{\NAT@aysep\NAT@spacechar}{\@citeb\@extra@b@citeb}%
     \NAT@date}}
  {\@citea\NAT@nmfmt{\NAT@nm}%
   \NAT@aysep\NAT@spacechar\NAT@hyper@{\NAT@date}}{}{}

\patchcmd{\NAT@citex}
  {\@citea\NAT@hyper@{%
     \NAT@nmfmt{\NAT@nm}%
     \hyper@natlinkbreak{\NAT@spacechar\NAT@@open\if*#1*\else#1\NAT@spacechar\fi}%
       {\@citeb\@extra@b@citeb}%
     \NAT@date}}
  {\@citea\NAT@nmfmt{\NAT@nm}%
   \NAT@spacechar\NAT@@open\if*#1*\else#1\NAT@spacechar\fi\NAT@hyper@{\NAT@date}}
  {}{}
\makeatother

\makeatletter
\DeclareRobustCommand{\lowcase}[1]{\@lowcase#1\@nil}
\def\@lowcase#1\@nil{\if\relax#1\relax\else\MakeLowercase{#1}\fi}
\makeatother

\DeclareSymbolFont{UPM}{U}{eur}{m}{n}
\DeclareMathSymbol{\umu}{0}{UPM}{"16}
\let\oldumu=\umu
\renewcommand\umu{\ifmmode\oldumu\else\math{\oldumu}\fi}

\if\papercls \othercls

\else

\fi

\let\oldsim=\sim
\renewcommand\sim{\ifmmode\oldsim\else\math{\oldsim}\fi}
\let\oldpm=\pm
\renewcommand\pm{\ifmmode\oldpm\else\math{\oldpm}\fi}
\newcommand\by{\ifmmode\times\else\math{\times}\fi}

\newcommand\tablebox[1]{\begin{tabular}[t]{@{}l@{}}#1\end{tabular}}
\newbox{\wdbox}
\renewcommand\c{\setbox\wdbox=\hbox{,}\hspace{\wd\wdbox}}
\renewcommand\i{\setbox\wdbox=\hbox{i}\hspace{\wd\wdbox}}

\newcount\timect
\newcount\hourct
\newcount\minct
\newcommand\now{\timect=\time \divide\timect by 60
         \hourct=\timect \multiply\hourct by 60
         \minct=\time \advance\minct by -\hourct
         \number\timect:\ifnum \minct < 10 0\fi\number\minct}

\catcode`@=11

\newcommand\comment[1]{}

\newcommand\commenton{\catcode`\%=14}

\renewcommand\math[1]{$#1$}
\newcommand\mathshifton{\catcode`\$=3}

\let\atab=&
\newcommand\atabon{\catcode`\&=4}

\let\oldmsp=\sp
\let\oldmsb=\sb
\def\sp#1{\ifmmode
           \oldmsp{#1}%
         \else\strut\raise.85ex\hbox{\scriptsize #1}\fi}
\def\sb#1{\ifmmode
           \oldmsb{#1}%
         \else\strut\raise-.54ex\hbox{\scriptsize #1}\fi}
\newbox\@sp
\newbox\@sb
\def\sbp#1#2{\ifmmode%
           \oldmsb{#1}\oldmsp{#2}%
         \else
           \setbox\@sb=\hbox{\sb{#1}}%
           \setbox\@sp=\hbox{\sp{#2}}%
           \rlap{\copy\@sb}\copy\@sp
           \ifdim \wd\@sb >\wd\@sp
             \hskip -\wd\@sp \hskip \wd\@sb
           \fi
        \fi}
\def\msp#1{\ifmmode
           \oldmsp{#1}
         \else \math{\oldmsp{#1}}\fi}
\def\msb#1{\ifmmode
           \oldmsb{#1}
         \else \math{\oldmsb{#1}}\fi}

\def\supon{\catcode`\^=7}

\def\subon{\catcode`\_=8}

\def\supsubon{\supon \subon}

\newcommand\actcharon{\catcode`\~=13}

\newcommand\paramon{\catcode`\#=6}

\comment{And now to turn us totally on and off...}

\newcommand\reservedcharson{ \commenton  \mathshifton  \atabon  \supsubon 
                             \actcharon  \paramon}

\catcode`@=12
\reservedcharson

\if\papercls \apjcls

\else

\fi

\if\papercls \othercls
\else
  \newcommand\inpress{n}
  \if\inpress y
    \received{\today}
    \revised{}
    \accepted{}
    \if\papercls \apjcls
    \slugcomment{}
    \fi
  \else
  \slugcomment{\tablebox{In preparation for {\em ApJ}. DRAFT of {\today}.}}
  \fi
\fi

\newcommand\chisq{\ifmmode{\chi\sp{2}}\else\math{\chi\sp{2}}\fi}
\newcommand\redchisq{\ifmmode{ \chi\sp{2}\sb{\rm red}}
                    \else\math{\chi\sp{2}\sb{\rm red}}\fi}
\newcommand\Teq{\ifmmode{T\sb{\rm eq}}\else$T$\sb{eq}\fi}
\newcommand\mjup{\ifmmode{M\sb{\rm Jup}}\else$M$\sb{Jup}\fi}
\newcommand\rjup{\ifmmode{R\sb{\rm Jup}}\else$R$\sb{Jup}\fi}
\newcommand\msun{\ifmmode{M\sb{\odot}}\else$M\sb{\odot}$\fi}
\newcommand\rsun{\ifmmode{R\sb{\odot}}\else$R\sb{\odot}$\fi}
\newcommand\mearth{\ifmmode{M\sb{\oplus}}\else$M\sb{\oplus}$\fi}
\newcommand\rearth{\ifmmode{R\sb{\oplus}}\else$R\sb{\oplus}$\fi}

\shorttitle{ApJ Template}
\shortauthors{Batta and Ramirez-Ruiz}

\begin{document}

\title{Accretion Feedback from newly-formed black holes and its implications for LIGO Sources}
\author{Aldo Batta\altaffilmark{1,2,3} and
Enrico Ramirez-Ruiz\altaffilmark{2,3}}

\affil{\sp{1} Instituto Nacional de Astrof\'isica, \'Optica y Electr\'onica, Tonantzintla, Puebla 72840, M\'exico: abatta@inaoep.mx\\
\sp{2} Department of Astronomy and Astrophysics, University of California, Santa Cruz, CA 95064, USA\\
\sp{3} Niels Bohr Institute, University of Copenhagen, Blegdamsvej 17, 2100 Copenhagen, Denmark}

\begin{abstract}Most common formation channels of stellar mass black hole (BH) binaries like the ones observed by LIGO, often assume they are assembled from the direct collapse of massive pre-supernova stars. However, it is still unclear whether the final mass and spin of the newly formed BH arises from the collapse of the entire stellar progenitor or just a fraction of it, given that coupling of accretion feedback released during BH formation to the surrounding infalling star will inevitably lead to its ejection. If the BH is built up via disk accretion, outflows from the center will result in residual gas ejection, thus halting the stellar collapse and reducing the amount of mass and spin that can be accreted by the newly formed BH. Here we discuss the general properties of BHs (mass and spin) associated with the collapse of rotating, helium star pre-supernova progenitors. When accretion feedback is included, the BH drives powerful outflows that heat the surrounding envelope, effectively shutting down the collapse. This gives rise to various outcomes ranging from very massive BHs with low spins, as inferred for GW150914, to lighter and faster-spinning BHs, as deduced for GW151226. 
\end{abstract}

\keywords{black holes, direct collapse, stellar evolution}


\section{INTRODUCTION}
\label{intro}
Our current, incomplete understanding of stellar mass black hole (BH) formation has been significantly  challenged by the detection of LIGO sources on the first two observing runs (GW150914, GW151226, GW170104, GW170608, GW170729, GW170809, GW170814, GW170818 and GW170823), resulting  from the mergers of binary BH systems (BHBs) with masses ranging from $7$ to $50 M_\odot$ and effective spins consistent with $\chi_{\rm eff}\lesssim 0.1$  \citep{GW150914_Abbott,GW151226_Abbott,GW170104_Abbott,GW170608_Abbott,GW170814_Abbott,LIGO_Catalog2018}. 

 LIGO observations of the mass-weighted angular momentum perpendicular to the orbital plane  $\chi_{\rm eff}$  have been argued to provide important constraints on the formation channel of BHBs \citep{Rodriguez_2016b,Farr_2017}. BHBs formed from field binary evolution are expected  to be aligned with the orbital angular momentum, unless the BHs receive a large natal kick as has been observed in some X-ray binaries \citep{Willems_2005,Fragos_2009,Mandel_2016b, Mirabel_2017}. This is true for the classical field formation channel  \citep{Voss_2003,Dominik_2012,Dominik_2013,Belczynski_2016} in which a wide massive binary undergoes a series of mass transfer episodes, usually involving a common envelope phase \citep[e.g.][]{Ivanova_2013,Ari2017} that finally leads to a tight BH binary, as well as for the chemically homogeneous evolution (CHE) channel \citep{deMink_2016,Mandel_2016,Marchant_2016}, in which a massive, close, tidally locked  binary evolves  chemically homogeneously and avoids a red giant phase due rapid rotation. Meanwhile, BHBs formed from the dynamical interaction of BHs and stars in a dense stellar system \citep{Sigurdsson_1993,Zwart_2000, Downing_2010,Downing_2011,Ziosi_2014, Samsing2014, Rodriguez_2015,Rodriguez_2016a,Rodriguez_2016c,Samsing2017,Samsing2018,Martin_2018} will yield BHs with roughly randomly aligned spins. 

On both formation channels, the assembly of BH binaries like GW150914, GW170104, GW170729, GW170809, GW170814, GW170818 and GW170823 may necessitate a {\it failed} supernova (SN) scenario in order to explain the large BH masses inferred ($\gtrsim30M_{\odot}$). The outcome of such  failed SNe could range from the quiet disappearance of a star yielding no observable transient to the production of  a highly luminous  transient \citep{Woosley_1993,Fryer2009, 2009ApJ...692..804L,Moriya2010,Lazzati_2012,2012ApJ...744..103M,Piro2013,Lovegrove2013,Dexter2013}.  If we were to venture on a general classification scheme for the outcome of failed supernova (SN) explosions, on the hypothesis that the central engine involves BHs, we would evidently expect  the rate at which the gas is supplied to the BH and  the angular momentum of the infalling material to be vital parameters. The amount and structure of fallback material is mainly determined  by the currently unknown properties of the explosion, which in turn depends on the debated internal structure of the stellar progenitor at the end of its life \citep{Perna2014}.  Yet, the angular momentum  is probably  the most crucial parameter \citep{2006ApJ...641..961L,2009MNRAS.398.2005Z}, as it
determines the geometry of the accretion flow. Even a little rotation can have a big impact, breaking  spherical symmetry and building accretion disks instead
of radial inflow.  

While spherical accretion onto BHs is inefficient, the dissipation rate can change dramatically if an accretion disk forms. In this case,  the flow onto the BH can liberate gravitational potential energy at a rate approaching a few tenths of $\dot{M} c^2$ \citep{Feng_Narayan_2014}, where $\dot{M}$ is the mass inflow rate. Therefore, even if only a small fraction of the energy generated  via accretion is effectively transported  into the infalling envelope, the collapse could be halted  \citep{Kohri_2005} and the mass and spin of the BH would differ significantly from that obtain by direct collapse of a non-rotating star.  Feedback energy from accretion onto stellar mass BHs is commonly observed  in Galactic X-ray binaries \citep{Ponti_2012} and the formation of a strongly magnetized wind outflow has been found to be a natural outcome of general relativistic, magnetohydrodynamical  simulations of accretion disks around BHs \citep{Tchekhovskoy_2011, McKinney_2012}.  The focal point of this paper is  the direct collapse of massive stars and the effects that the ensuing  feedback, which arises when an accretion disk is formed, has on the properties of the newly formed BH.  When stellar  rotation  is taken into account, it is argued that the final properties of stellar mass BHs that formed in these failed SN explosions can be significantly different  from those derived assuming the gas has no angular momentum.


\section{Formation of BHs from direct collapse}

Due to the complexities associated with angular momentum transport inside massive stars, it is difficult to make detailed predictions about the rotational profile  of pre-SN stellar progenitors. Differentially rotating stars tend to move towards rigid rotation when acted upon by processes which transport angular momentum, such as magnetic coupling or torques \citep{Zahn_1975,Spruit_2002}. However, such mechanisms are not ubiquitous, and mass loss via winds, convection and mass transfer in binary systems can also play an important role in shaping the angular momentum distribution of massive stars. Regardless these complications, given a pre-SN stellar progenitor with an angular velocity distribution $\Omega(r)$, one can obtain the growth of a BH formed from direct collapse, as shells of material (represented by a spherically symmetric density profile) fall back onto the newly formed BH. 
\begin{figure}[!h]
\begin{center}
 \includegraphics[trim={1.2cm 2.6cm -0.4cm 0.4cm},height=0.49\textwidth,angle=-90]{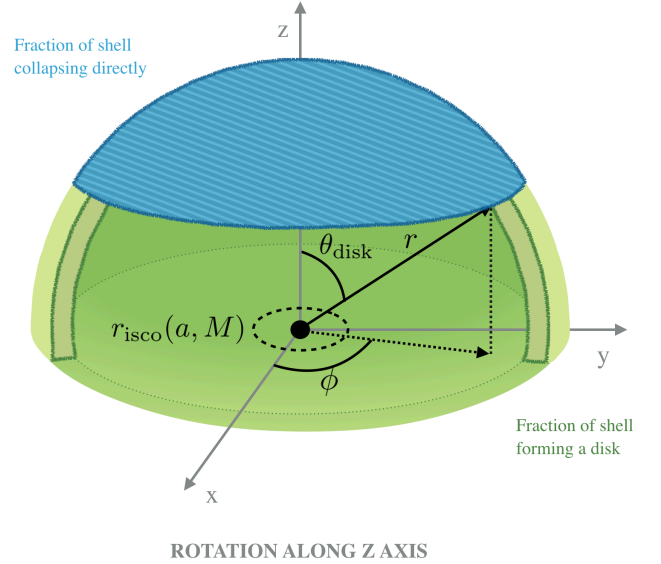}
      \caption{Top hemisphere of stellar shell about to collapse from a radius $r$ onto a BH with mass $M$ and spin $a$. Material located at $\pi/2\geq\theta\geq\theta_{\rm disk}$ (green colored) has enough angular momentum to form a disk around the BH ($j(r,\theta)\geq j_{\rm isco}$). Material located at $\theta<\theta_{\rm disk}$ (blue colored with stripes) has little angular momentum ($j(r,\theta)< j_{\rm isco}$) and will collapse directly onto the BH.}

 \label{fig:Shell_disk}

 \end{center}
 \end{figure}

\subsection{Initial BH's Mass and Spin}
 As we will demonstrate below, the newly formed BH might be unable to accrete all of the available stellar material. In order to calculate the final mass and spin of the BH resulting from direct collapse, we need to consistently follow its accretion history soon after it is formed. In what follows, we assume that a BH forms after innermost $M_{\rm core} = 3M_{\odot}$ has collapsed. We find the exact value of this assumption has little impact to the validity of our conclusions. Stellar material with mass coordinates  $\gtrsim M_{\rm core}$ is then assumed to have a symmetric rotational profile along the $z$ axis, whose angular velocity distribution  is given by $\Omega(r)$, where $r$ is the spherical radius (Figure~\ref{fig:Shell_disk}). The angular momentum content of the initial BH with  mass $M_{\rm core}$ is given by
\begin{equation}
J_{\rm core} =   \int_{M_{\rm core}} \Omega(r) \ r^{2} \sin^2\theta\ dm,
\label{eq:AngMom}
\end{equation}
where $\theta$ and $\phi$ are the polar and azimuth angles, respectively, and $dm = \rho\ r^2 \sin \theta\ dr\ d\theta\ d\phi$. 

By making use of the mass and angular velocity distributions derived from stellar evolution models we can robustly calculate the angular momentum content of the initial $3M_{\odot}$ BH. The BH's initial spin parameter can be written as 
\begin{equation}
a_{\rm core} = \frac{J_{\rm core}c}{GM_{\rm core}^2}.
\end{equation}
Once the initial BH's mass and spin parameter are known, the location of the innermost stable circular orbit (ISCO) $r_{\rm isco}$ can be calculated, which in turn determines  the specific angular momentum needed for infalling material to form an accretion disk $j_{\rm isco}$. The properties of the ISCO around a Kerr BH on the equatorial plane were obtained in a seminal paper by \cite{Bardeen_1972}. The radius at the ISCO, scaled by $GM_{\rm bh}/c^2$, can be written as:
\begin{equation}
r_{\rm isco} = 3 + z_2 \pm \left[(3 - z_1)(3 + z_1 + 2 z_2) \right]^{1/2} \ ,
\label{eq:risco}
\end{equation}
where $z_1$ and $z_2$ are determined by the BH's spin according to:
\begin{equation}
\begin{split}
z_1 &= 1 +  \left(1 - a^2\right)^{1/3}  \left[ (1 + a)^{1/3} + (1 - a)^{1/3} \right] \\
z_2 &= (3  a^2 + z_1^2)^{1/2}\ .
\end{split}
\end{equation}
As is evident in equation (\ref{eq:risco}),  the location of $r_{\rm isco}$ has two solutions, which correspond to prograde and retrograde equatorial orbits around the BH. For retrograde orbits, the third term in equation (\ref{eq:risco}) is positive, yielding a radius that increases with the  spin of the BH. For prograde orbits characterized by an angular momentum aligned with the spin of the BH, as in our direct collapse scenario, the third term in equation (\ref{eq:risco}) is negative, yielding a radius that decreases with $a$. Since the location of $r_{\rm isco}$ determines the angular momentum needed to orbit the BH at the ISCO $j_{\rm isco}$, a larger amount of angular momentum is required for retrograde orbits than for prograde orbits. The specific angular momentum at the ISCO, scaled by $GM_{\rm bh}/c$, can be written as:
\begin{equation}
j_{\rm isco} = \frac{2}{3^{3/2}}\left[ 1 +2(3r_{\rm isco} - 2)^{1/2}  \right]
\end{equation}
and provides a limit for the amount of specific angular momentum that can be carried by material accreted by the BH through an accretion disk.

\subsection{Calculating the properties of newly-formed BHs}
\label{sec:BHproperties}
In our formalism, a collapsing shell of mass $m_{\rm shell}$ located at a radius $r$ has an specific angular momentum distribution that can be described by:
\begin{equation}
j(r,\theta) = \Omega(r)\ r^2 \sin^2\theta .
\label{eq:specj}
\end{equation}
Since only material with $j(r,\theta) \geq j_{\rm isco}$ will form an accretion disk, it is useful to define the disk formation angle:
\begin{equation}
 \theta_{\rm disk} = \arcsin\left[\left(\frac{j_{\rm isco}}{\Omega(r) r^2}\right)^{1/2}\right],
 \label{eq:thetaC}
\end{equation}
which is the critical polar angle separating the directly infalling and the centrifugally supported material. The diagram in Figure \ref{fig:Shell_disk} shows the portion of a shell that forms an accretion disk (green region towards the equator with $\theta_{\rm disk}\leq\theta\leq\pi/2$) and the portion that collapses directly onto the BH due to its low angular momentum (blue region towards the pole with $\theta<\theta_{\rm disk}$). Each spherical shell is assumed to collapse in a dynamical time scale $t_{\rm dyn} \simeq(r^3/GM)^{1/2}$ and will be accreted either directly (when $j<j_{\rm isco}$) or through an accretion disk (when $j\geq j_{\rm isco}$) in a viscous time scale $t_{\nu} \approx \alpha^{-1}(H/R)^{-2} t_{\rm orb}$. Here $H$ is the disk's scale height,  $\alpha$ is the viscosity parameter and  $t_{\rm orb}$ the Keplerian orbital period at an accretion radius $R$. 

Thus, assuming that the disk formed from the collapse of a shell is accreted before the next shell collapses, we can evolve the BH's mass and spin as it accretes material with $j\geq j_{\rm isco}$ through a thin accretion disk, and material with $j< j_{\rm isco}$ in a quasi-radial flow. For this to be the case, the viscous timescale at the circularization disk radius $R_{\rm circ}$ should be shorter than the fallback time of the following shell, which we generally found to be a defensible assumption and will be further discussed in the last section of this paper. 

According to \cite{Bardeen_1970} and \cite{Thorne1974}, the mass accreted through a thin disk will transfer  angular momentum and mass to the BH given by:
\begin{equation}
\begin{split}
\Delta J_{\rm disk} &= j_{\rm isco}m_{\rm disk}\\
\Delta M_{\rm disk} &= e\ m_{\rm disk}\ , 
\end{split}
\label{eq:deltaJM}
\end{equation}
where $m_{\rm disk}=(1-\cos\theta_{\rm disk})m_{\rm shell}$ is the mass forming an accretion disk, and $e = (1- 2/3r_{\rm isco})^{1/2}$ is the orbital binding energy for a particle at the ISCO, which must be lost (radiated) in order to fall to the BH and get accreted. This translates into an angular momentum contribution from the entire shell given by: 
\begin{equation}
\begin{split}
J_{\rm shell}& = 2\int_{\theta < \theta_{\rm disk}} \Omega(r) \sin^{3}\theta \ \rho\ r^4 \ dr\ d\theta\ d\phi \\
 &+ 2\int_{\theta \geq \theta_{\rm disk}} j_{\rm isco}\ \rho\ r^2 \sin\theta\ dr\ d\theta\ d\phi,
\end{split}
\label{eq:Jshell}
\end{equation}
where the factor 2 comes from integrating over the two hemispheres. The first term in equation (\ref{eq:Jshell}) represents material with low angular momentum, that will collapse directly to the BH, transferring its entire mass and angular momentum. The second term in equation (\ref{eq:Jshell}), on the other hand, corresponds to material that will form an accretion disk, transferring only the specific angular momentum at the ISCO.  Thus, following the orderly collapse of the spherical shells of mass $m_{\rm shell}$ within the star, we can evolve the BH as it accretes each of the infalling shells according to their mass and angular momentum, where the mass coordinate $M/M_{\rm star}$ will indicate the fraction of the star that has collapsed into the BH. 

\subsection{BH formation from the direct collapse of rapidly rotating pre-SN He stars}
\label{sec:nofeed}
As an example of the simple method outlined here, we make use of the four most massive (at Zero Age Main Sequence) and rapidly rotating models of long GRB progenitors from \cite{Woosley2006} (WH06). These collapsing stars, with the same initial mass $M_{\rm i} =35M_{\odot}$ and angular momentum $J_{\rm i} = 1.4\times10^{53} \text{erg s}^{-1}$, have different final masses ($\approx 34M_{\odot}$, $\approx28M_{\odot}$, $\approx21M_{\odot}$ and $\approx13M_{\odot}$) as the result of different mass loss prescriptions (due to winds and magnetic torques). Thus, it is expected that models with the smaller total mass loss, are the ones with the larger angular momentum content. 

\begin{figure}[t]
\begin{center}
 \begin{minipage}[c]{0.49\textwidth}
\includegraphics[trim={1.3cm 0.2cm 1.5cm 1.2cm},clip,width=\textwidth]{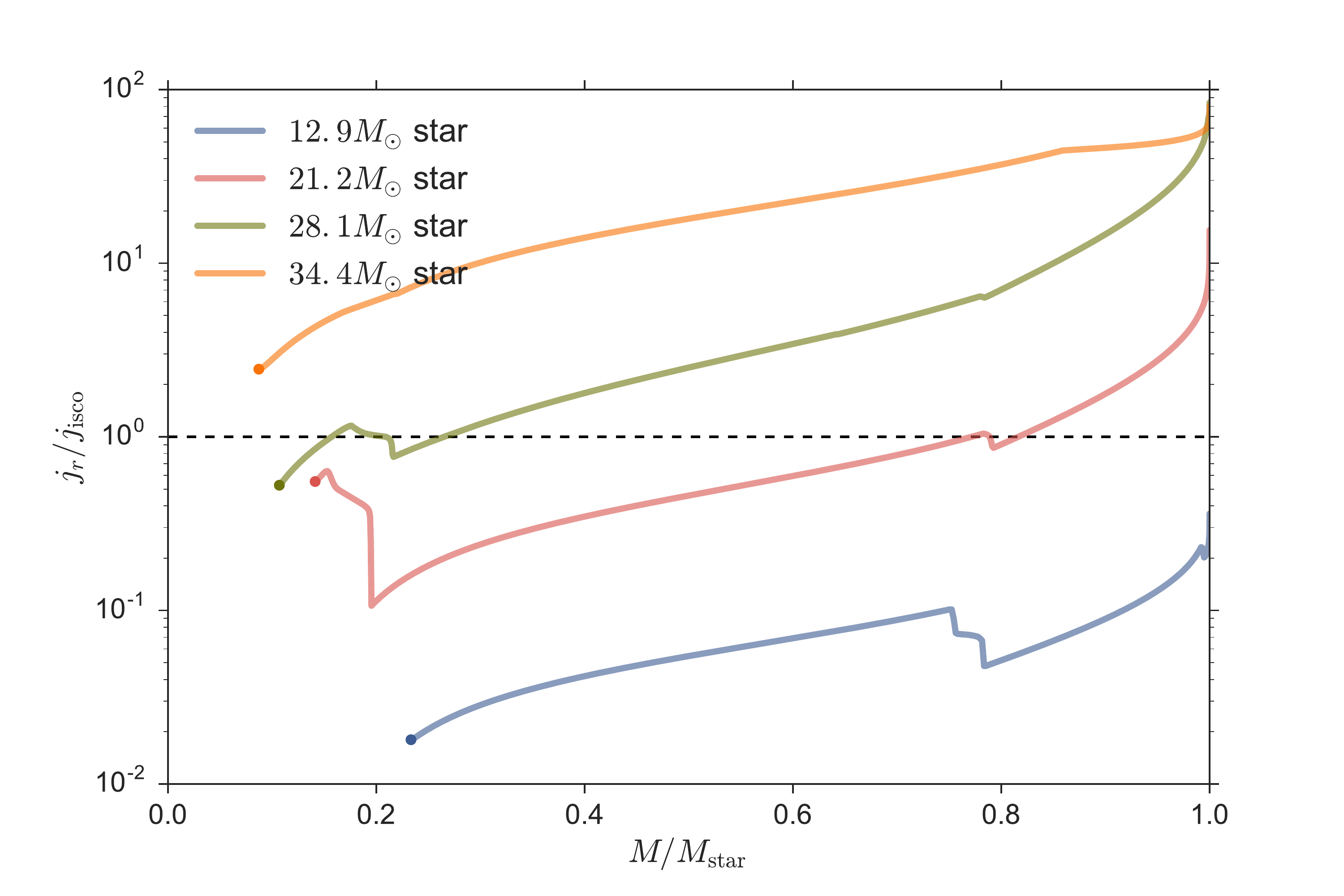}
\end{minipage}

 \begin{minipage}[c]{0.49\textwidth}
 \includegraphics[trim={1.3cm 0.2cm 1.5cm 1.2cm},clip,width=\textwidth]{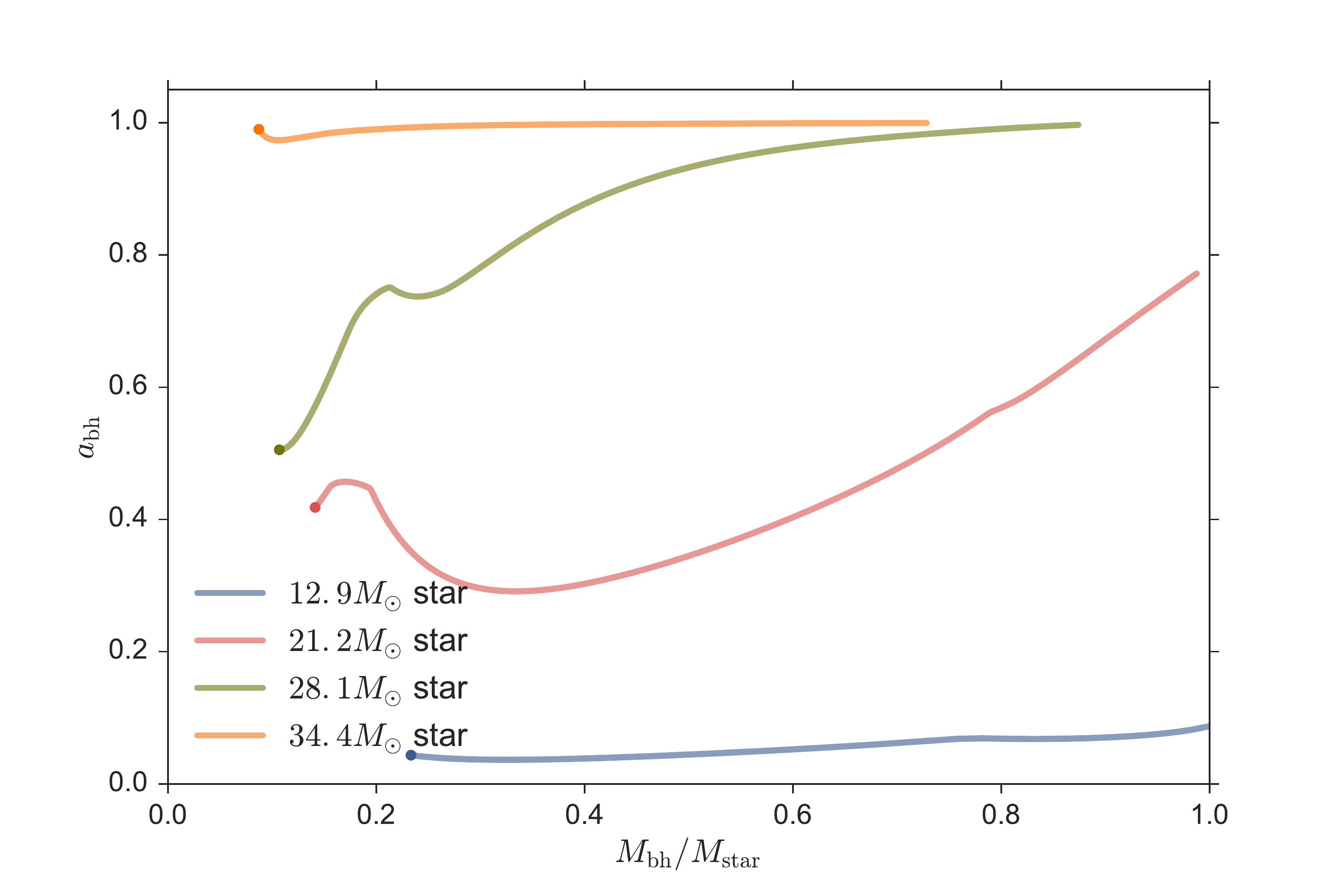}
\end{minipage}

\begin{minipage}[c]{0.49\textwidth}
 \includegraphics[trim={1.3cm 0.2cm 1.5cm 1.2cm},clip,width=\textwidth]{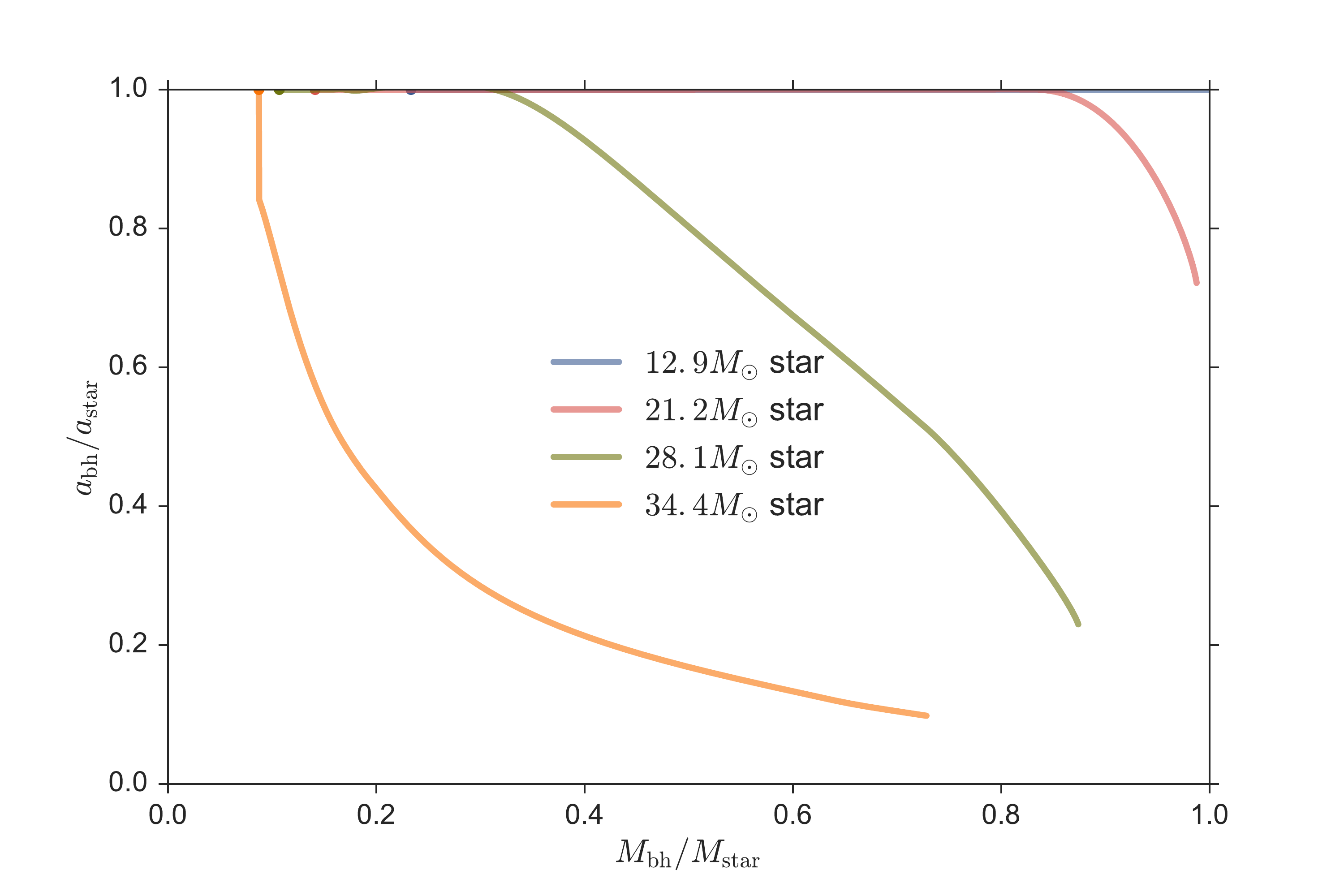}
\end{minipage} 

\caption{Top panel: Ratio $j/j_{\rm isco}$ at the equator for WH06 models with $34M_{\odot}$, $28M_{\odot}$, $21M_{\odot}$ and $13M_{\odot}$ (orange, green, red and blue lines respectively).  Middle panel: Evolution of the BH's spin parameter $a_{\rm bh}$ as function of the BH's mass. Bottom panel: Evolution of the ratio between the BH's spin parameter $a_{\rm bh}$ and the star's spin parameter $a_{\rm star}$ as function of the BH's mass. The points on the left of the curves correspond to the location of the initial BH mass of $3M_{\odot}$. We note that these models do not limit the rotation of the star by taking into account a centrifugal force term, which affects the most massive model, whose outermost $\approx 10\%$ mass rotates above breakup. In this case we limit the angular velocity to be $\Omega(r)=0.95\ \Omega_{\rm break}(r)$ when $\Omega(r)>0.95\ \Omega_{\rm break}(r)$. }

\label{fig:JDist}
\end{center}
\end{figure}

The top panel from Figure \ref{fig:JDist} shows the evolution of the ratio between the star's specific angular momentum and the angular momentum needed to form an accretion disk $\beta = j/j_{\rm isco}$, as a function of the normalized mass coordinate $M/M_{\rm star}$, for the selected stellar models.  $j_{\rm isco}$ changes as the BH's mass and spin parameter ($M_{\rm bh}$, and $a_{\rm bh}$ respectively) evolve with the accretion of mass and angular momentum from the infalling shells, as illustrated in the middle panel of Figure \ref{fig:JDist}.  Note that since only a fraction of the mass is accreted by the BH (equation \ref{eq:deltaJM}), all stars with enough angular momentum to form an accretion disk will yield BHs with masses that are only a fraction of the initial star's mass. Moreover, since the BH can only accrete material with $j\leq j_{\rm isco}$, the amount of angular momentum accreted by the BH will be smaller than the one contained in the stellar material with $j>j_{\rm isco}$, resulting in a spin $a_{\rm bh}$ that is smaller than that of the entire star $a_{\rm star} = c J_{\rm star}/(GM_{\rm star}^2)$. This can be seen in the bottom panel of Figure \ref{fig:JDist}. As expected, only stars rotating rapidly enough to form an accretion disk produce BHs with different mass and spin from their progenitor stars. Given that the BH is accreting both low and high angular momentum material from each collapsing shell (see Figure \ref{fig:Shell_disk}), it can take a large fraction of the star with $j>j_{\rm isco}$ to increase the BH's spin to high values ($\gtrsim0.8$). 

Within this simple formalism, the final mass and spin of the BH is solely determined by the initial mass and angular momentum distribution  of the star. The final mass of the BH differs from that of the star due to radiation of the orbital energy at the ISCO. This evolution can be altered drastically if we consider the scenario where a hot, magnetized accretion disk produces an outflow that injects mechanical energy onto the infalling layers. It is to this issue that we now turn our attention. 

\section{Effects of feedback on BH formation}
\label{sec:feed}

\begin{figure}[h!]
\includegraphics[trim={1.2cm 0.5cm 1.2cm 0.8cm},clip,width=0.49\textwidth]{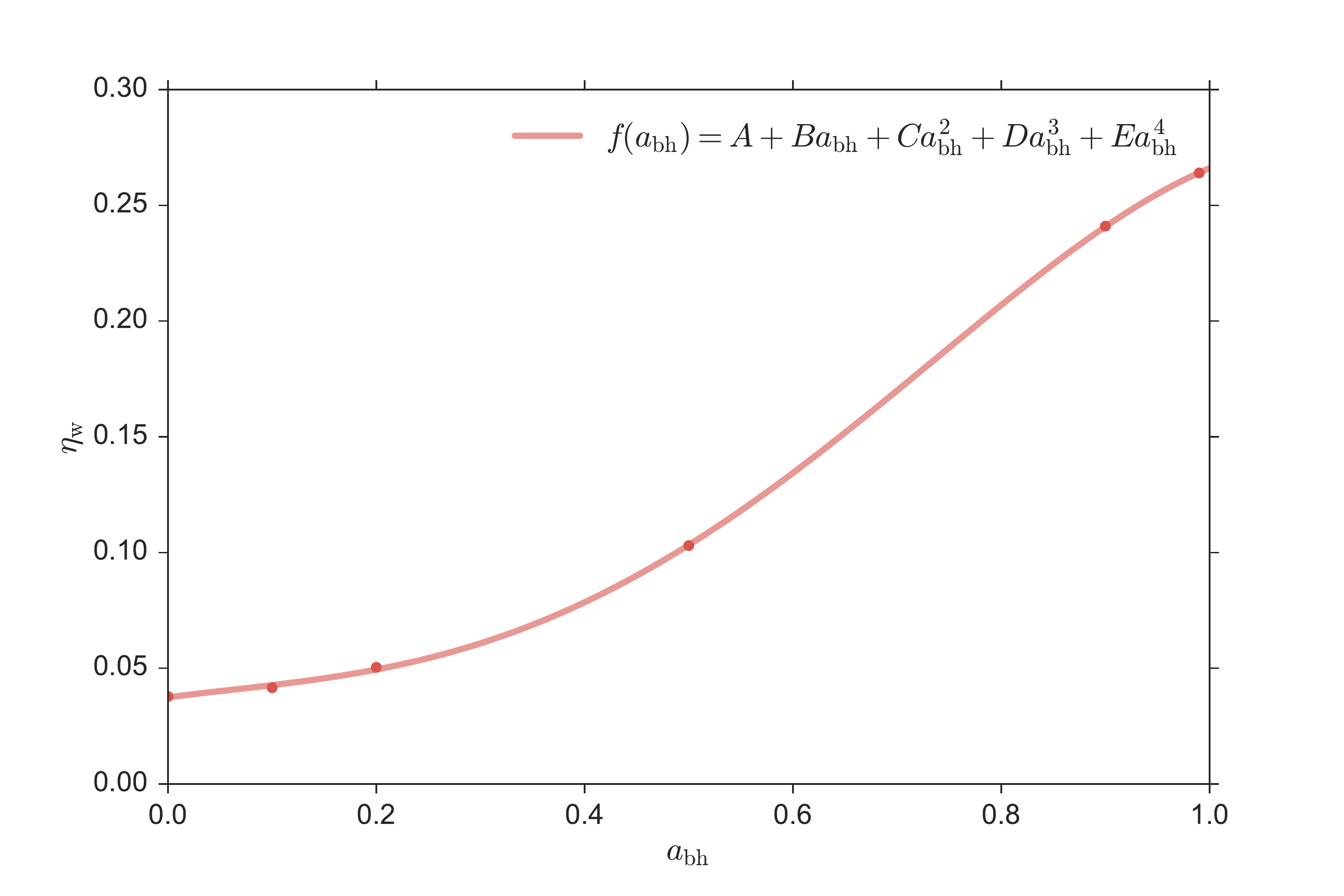}
\caption{Outflowing wind efficiency derive by McK12 as a function of the spin of the BH. A polynomial fit is shown on top of the data points from Table 5 in McK12 where $A=0.037, B=0.59, C=-0.137 , D=0.826, E=-0.519$.}
\label{fig:Feedback_McK12}
\end{figure}

As has been shown in, for example, \citet{McKinney_2012} (McK12), MHD GR simulations of accretion disks around rotating BHs generally show the formation of a magnetized wind outflow, which can inject energy that can be potentially shared with the infalling layers of the star. From Table 5 in McK12 one can obtain the wind efficiency $\eta_{\rm w}$ as a function of the BH spin, which is shown in Figure \ref{fig:Feedback_McK12}.  
This wind efficiency represents the fraction of the mass accretion rate, $\dot{M}_{\rm bh}$, that is ejected. This parametrization does not include a value for the kinetic energy of the wind. In what follows we introduce an efficiency parameter, $\epsilon$, that describes the total kinetic  energy content of the flow. That is, $\dot{E}_{\rm wind}=\eta_{\rm w}\epsilon \dot{M}_{\rm bh} c^2$. Here $\epsilon=(v_{\rm wind}/c)^2$, where $v_{\rm wind}$ is the characteristic velocity of the wind, and $\eta_{\rm w}=\dot{M}_{\rm wind}/\dot{M}_{\rm bh}$.

As an inner shell with mass $m_{\rm shell}$ collapses onto the BH in a dynamical time scale, the adjacent outer shells will continue to their radial infall. As a result,  the kinetic and binding energies of the shells will be altered. The kinetic energy gained from the collapse makes it more difficult to unbind the star, since now the infalling shells have to be stalled before they can be pushed out by BH feedback. Furthermore, the shell's binding energy also increases as the shells move to smaller radii. In order to take this properly into account, we  solve the equation of motion (EoM) of the infalling shells in order to obtain their kinetic and potential energy as they collapse. This can be easily done (when the internal pressure of the star is neglected) by writing the EoM of a shell located at a given radius $r$, whose motion is solely determined by the mass interior to it. 

When a single accreted mass shell can produce enough feedback energy to unbind the outer layers of the star ($E_{\rm fb, shell}\geq  E_{\rm ub}$), we assume that the BH has reached its final mass and spin. However, if the energy produced by a single shell is not able to unbind the star ($E_{\rm fb, shell}< E_{\rm ub}$) we need to define a prescription for how  this energy is accumulated. To do this, we apply two different prescriptions in this paper. 

In the first prescription, which we named {\it integrated feedback without losses}, we assume that every time $E_{\rm fb, shell}<E_{\rm ub}$, the feedback energy accumulates with every shell that is able to form a disk: $E_{\rm fb} = \sum E_{\rm fb, shell}$. Eventually, the integrated feedback energy $E_{\rm fb}$ will increase beyond $E_{\rm ub}$ and unbind the star.  This provides an upper limit of the total energy available to unbind the star, since we do not account for the possibility of energy being advected.

In the second prescription, which we named  {\it integrated feedback with losses},  we assume that the feedback energy produced by the accretion of a shell is deposited fruitlessly into the infalling outer layers and the star will continue its collapse. This energy will be transformed (at least partially) into heat which can be advected and lost into the BH. In this model we assume that the efficiency of energy conversion is small when $E_{\rm fb,shell}<E_{\rm ub}$ and that the fractional contribution  of every shell that forms an accretion disk  to the total energy budget is $\psi=(E_{\rm fb,shell}/E_{\rm ub})$. The integrated feedback energy is then calculated by integrating the energy contribution of all the shells that are able to form an accretion disk: $E_{\rm fb} = \sum \psi E_{\rm fb, shell}$. This model represents a conservative lower limit to the total available energy to unbind the star. \\

\begin{figure}[!h]
\includegraphics[trim={1.2cm 0.5cm 1.2cm 0.8cm},clip,width=0.49\textwidth]{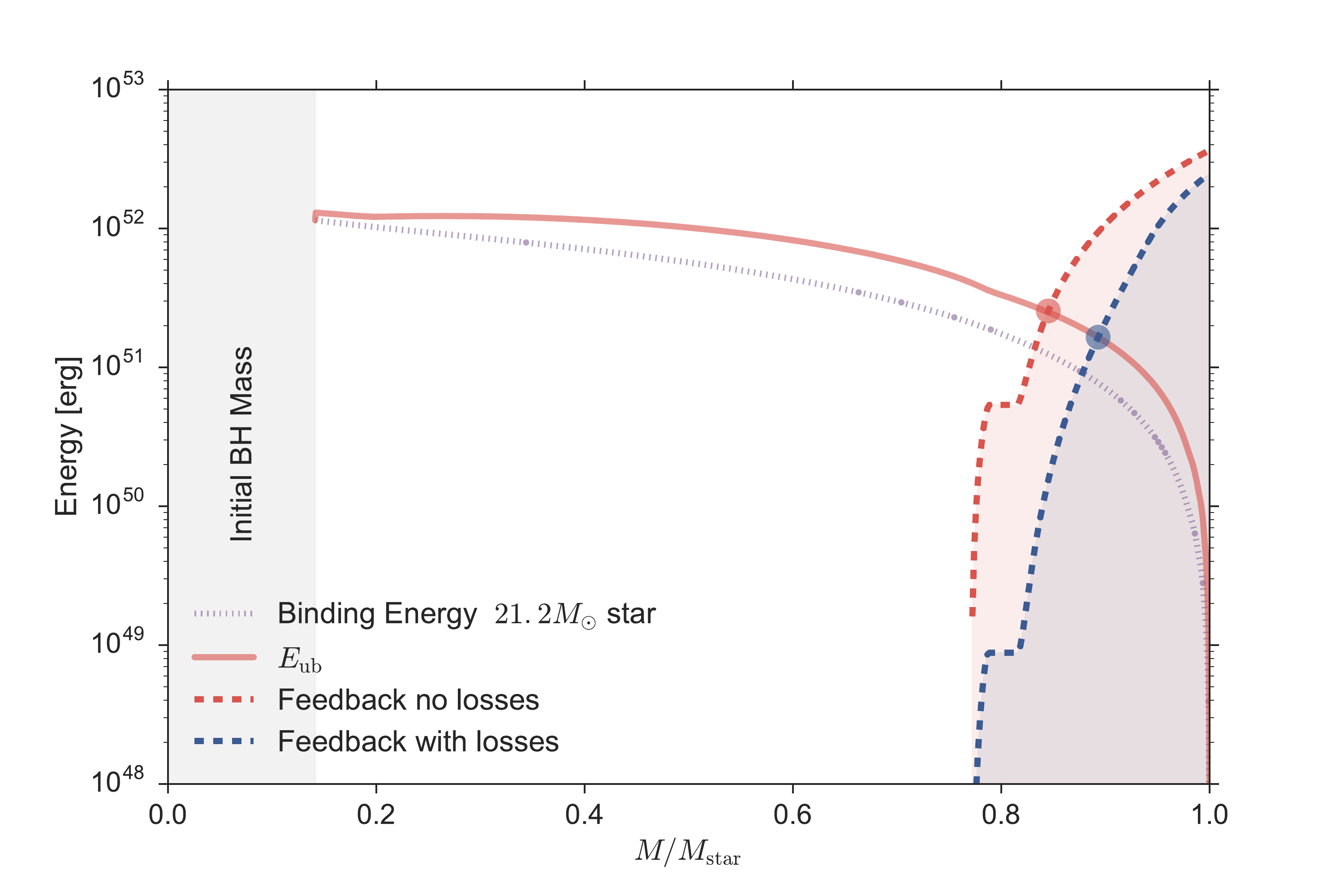}
\includegraphics[trim={1.2cm 0.5cm 1.2cm 0.8cm},clip,width=0.49\textwidth]{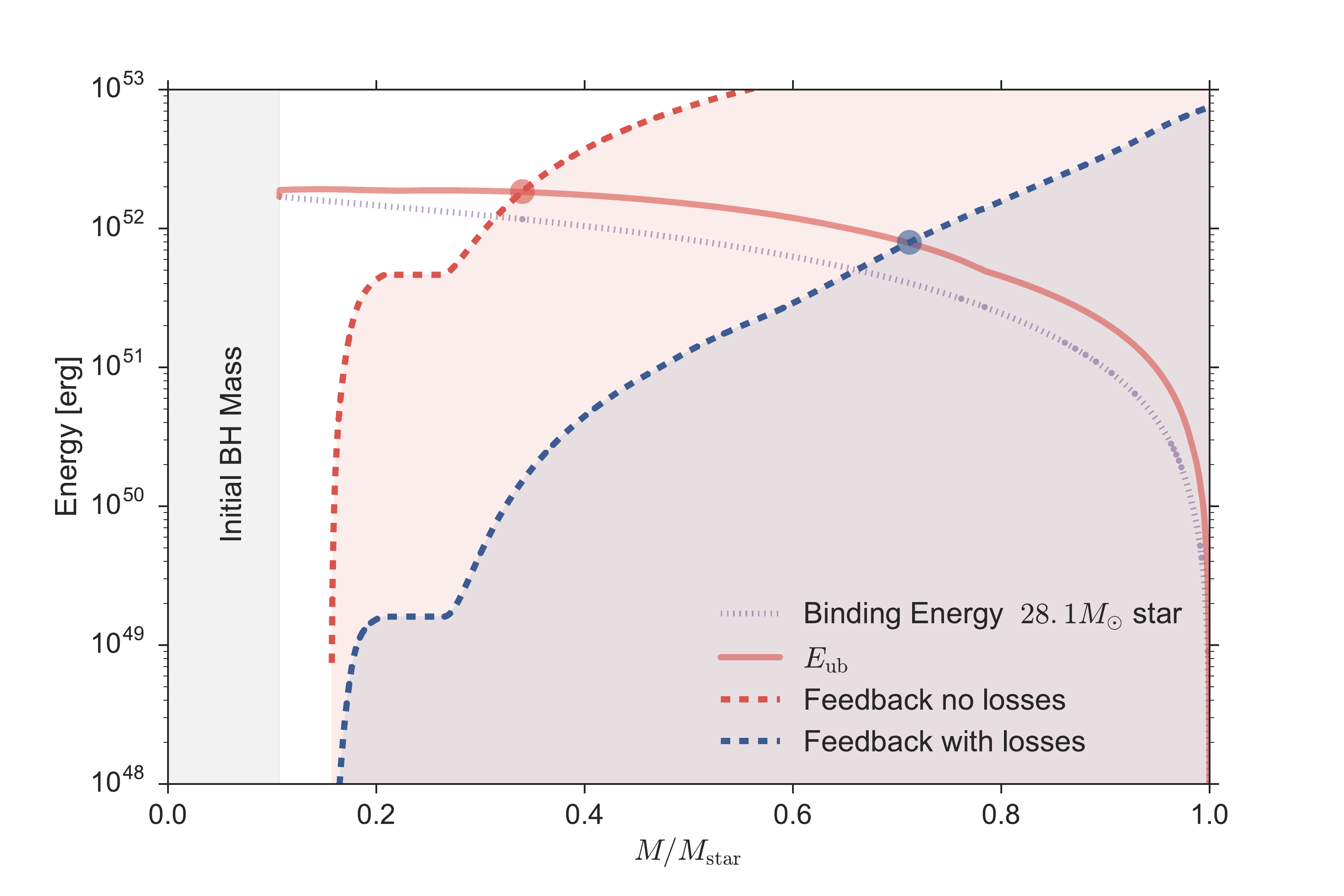}

\caption{Integrated feedback energy $E_{\rm fb}$ from the simple feedback ({\it red dashed}) and from the feedback with losses ({\it blue dashed line}) for the $21M_{\odot}$ and $28M_{\odot}$ WH06 stellar models. The {\it thick solid lines} correspond to the energy needed to unbind the collapsing shells $E_{\rm ub}$ outside the mass coordinate $M/M_{\rm star}$ while the {\it dotted lines} to the initial binding energy of the star. The {\it red and blue symbols} (for feedback with no losses and with losses, respectively) indicate the mass coordinate at which the integrated feedback energy $E_{\rm fb}$ matches the energy needed to unbind the collapsing shells $E_{\rm ub}$. The grey shaded area on the left shows the assumed initial BH mass. In both cases we use $\epsilon=10^{-3}$.}
\label{fig:Feedback_prescription}
\end{figure}

By applying the feedback prescriptions defined above to the WH06 stellar models we can determine the effects of feedback on the final  properties of the newly formed BH. Figure \ref{fig:Feedback_prescription} shows the initial binding energy ({\it dotted line}), the energy needed to unbind the collapsing star $E_{\rm ub}$ ({\it red solid line}), the integrated feedback energy $E_{\rm fb}$ assuming the feedback without losses ({\it red dashed line}), and the integrated feedback with losses ({\it blue dashed line}) for the $21M_{\odot}$ ({\it top panel}) and the $28M_{\odot}$ star ({\it bottom panel}). The red and blue points in Figure \ref{fig:Feedback_prescription} (for the feedback without and with losses respectively) show the mass scale at which the feedback energy $E_{\rm fb}$ matches the energy needed to unbind the collapsing star $E_{\rm ub}$. Compared to the feedback without losses, the feedback with losses allows a larger fraction of the star to collapse before significant energy feedback is injected. As expected, this translates into a higher BH mass and spin, since the specific angular momentum in the stellar models increases with the mass coordinate. 

\begin{figure}
\includegraphics[trim={1.6cm 0.5cm 1.2cm 0.8cm},clip,width=0.49\textwidth]{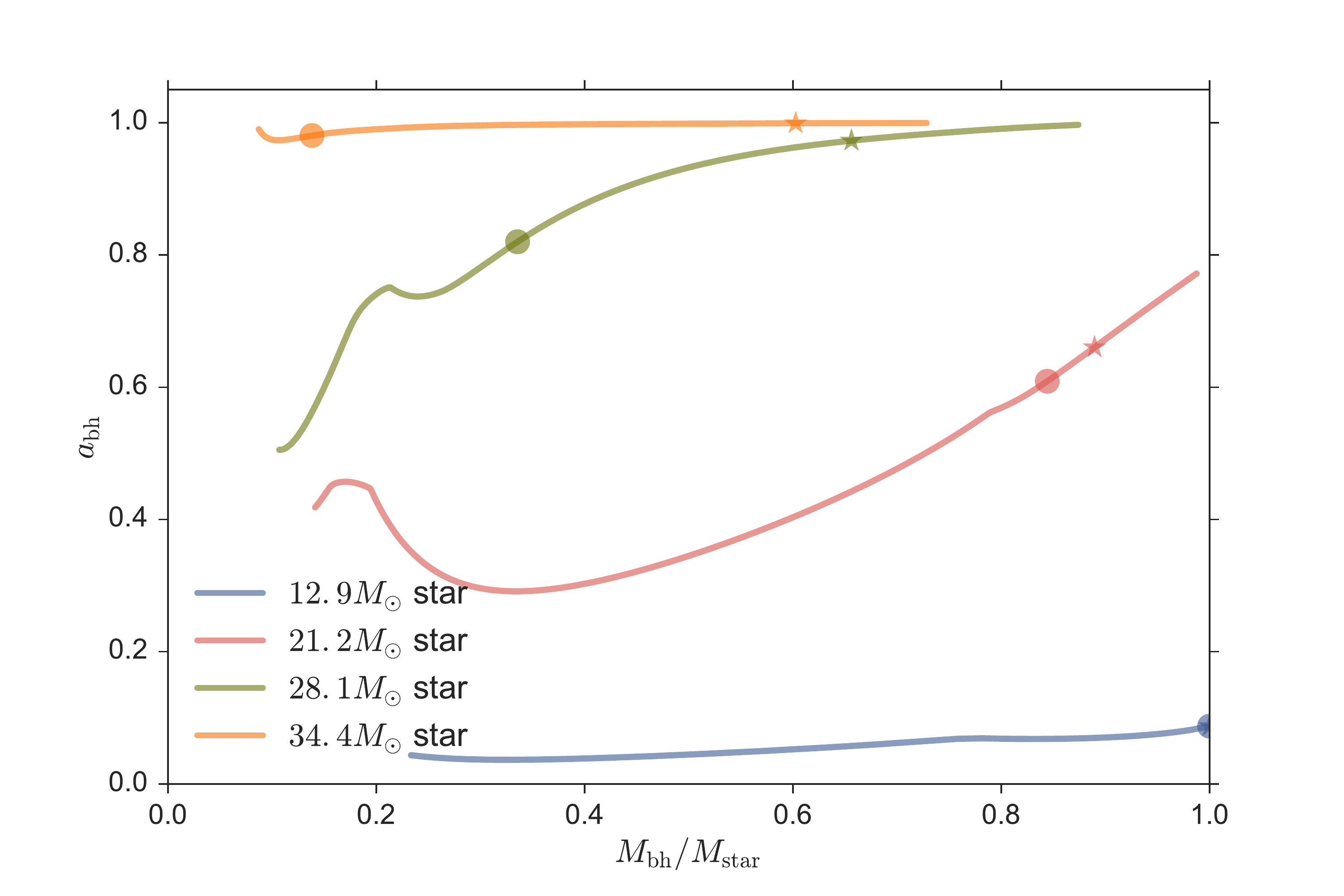}
\includegraphics[trim={1.4cm 0.0cm 1.2cm 0.8cm},clip,width=0.49\textwidth]{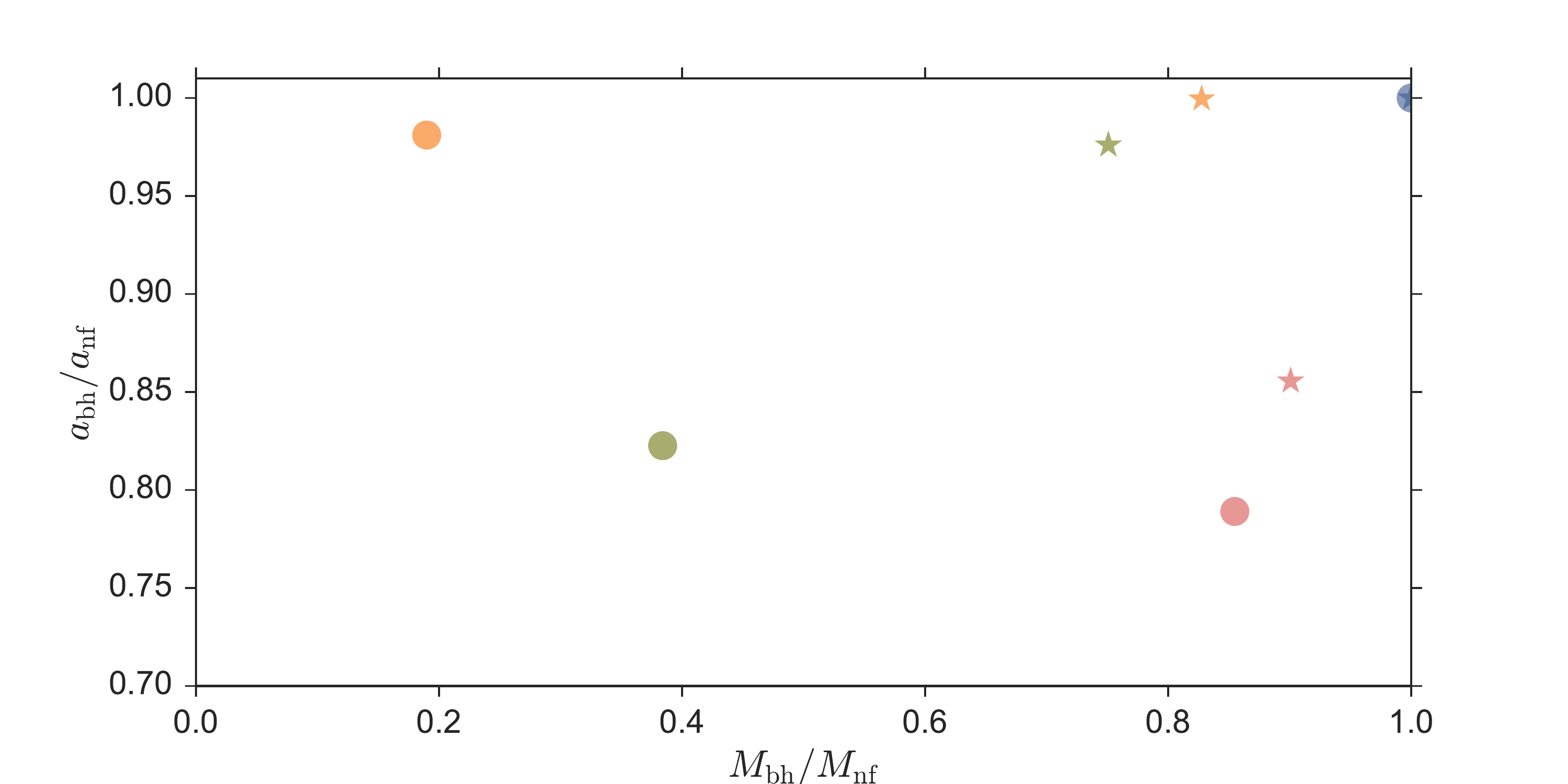}

\caption{Top panel: The same as the middle panel in Figure \ref{fig:JDist}. The circles indicate the  mass and spin of the BHs obtained when including feedback without losses while the star symbols shows the results  for the feedback with losses. In both cases we use $\epsilon=10^{-3}$. Bottom panel: Shows the fractional difference between models, with $(a_{\rm nf},M_{\rm nf})$ showing the expected parameters obtained assuming direct collapse without feedback.  The results using our two feedback prescriptions are denoted as $(a_{\rm bh},M_{\rm bh})$  with {\it dot} and {\it star} symbols  correspond to models with feedback without and with losses, respectively.  The only BH that has the same mass and spin in all scenarios is the one that is unable to form an accretion disk ({\it blue points}).}
\label{fig:Fb_MassSpin}
\end{figure}

Figure \ref{fig:Fb_MassSpin} shows the  mass and spin of the BHs obtained for WH06 models assuming feedback without losses ({\it circles}) and a feedback with losses ({\it stars}).
Also shown in the bottom panel is the fractional difference between models. Here $(a_{\rm nf},M _{\rm nf})$ are the final parameters of the BH assuming direct collapse without feedback, and $(a_{\rm bh},M_{\rm bh})$  are the values obtained using  the two different feedback prescriptions ({\it dots and star symbols}, respectively). The $13M_{\odot}$ star is the only one not forming an accretion disk and the BH's spin and mass is the same in all scenarios. However, BHs from rapidly rotating stars can have up to $\gtrsim20\%$ less spin and $\gtrsim80\%$ less mass that the one they would have without feedback. 

\begin{figure}[!h]
\includegraphics[trim={0.8cm 0.0cm 0.0cm 0.0cm},clip,width=0.49
\textwidth]{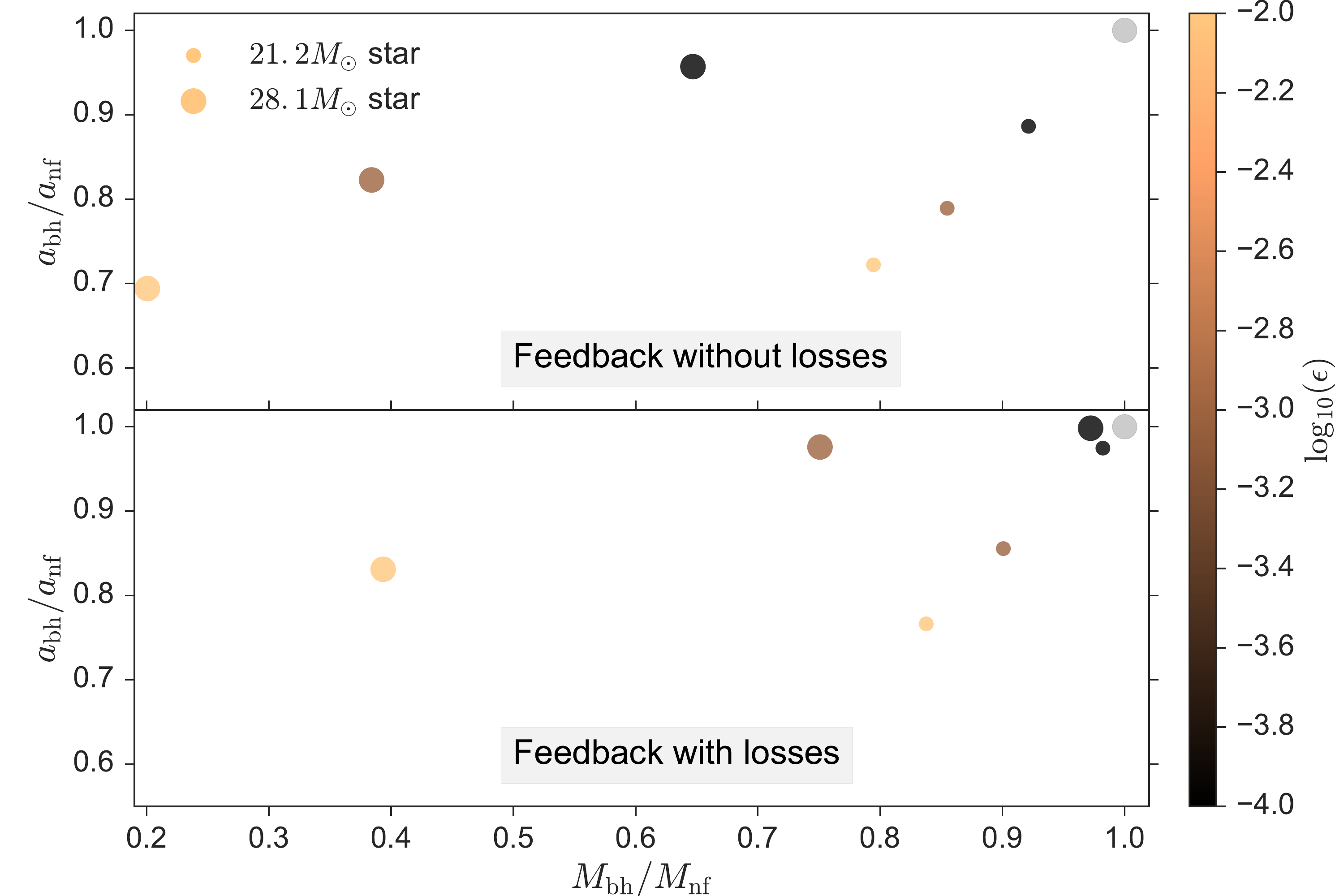}
\caption{Assessing the influence of $\epsilon$. Shown is the fractional difference between models when $\epsilon$ is altered. All model results are normalized to the results assuming direct collapse without feedback: $(a_{\rm nf},M_{\rm nf})$.  Here we calculate models using  $\epsilon=10^{-4}$, $10^{-3}$ and $10^{-2}$  (indicated by the color of the symbols) and for our two  feedback prescriptions ({\it top} and {\it bottom} panels). The size of the symbols indicate the mass of the model used, as indicated in the top panel.}
\label{fig:SpinMass_epsilon}
\end{figure}

Undoubtedly, our feedback prescriptions should be used with caution given their simplicity yet they clearly illustrate the profound effects that feedback can have in shaping the resultant  properties of newly formed BHs. As shown in Figure \ref{fig:Fb_MassSpin}, feedback prevents the formation of massive, rapidly spinning BHs. Thus, if feedback is efficient, the most massive BHs observed by LIGO must come from slowly rotating stars with little mass and spin loss, or from even heavier rapidly rotating progenitors with substantial mass and spin loss. This conclusion holds true when the efficiency of the disk wind is modified, as illustrated in  Figure \ref{fig:SpinMass_epsilon}. In the sections that follow we will use the simple method described above to further explore the importance of the star's angular momentum content in shaping the final BH mass and spin obtained from the collapse of these stars. 

\section{The Role of Stellar Rotation}
Given the uncertainties of angular momentum transport in massive stars, it is tempting to try use constraints from LIGO BHs to discern the properties of their progenitor systems.  Motivated by this, in this section we explore the  repercussions of stellar rotation  on the properties of newly formed black holes. As we have done in the previous section, we make use of the $21$ and $28M_{\odot}$ WH06 stellar evolution profiles but assume that they rotate as a rigid body. In this case, the angular velocity can be parametrized as $\Omega = \zeta\ \Omega_{\rm break}$, where $\zeta < 1$. This is a reasonable approximation for WH06 models, which despite being differentially rotating, have near constant angular velocity throughout a large fraction of the star. Some mild deviations are found in the outer-layers, which contain a small fraction of the mass, and in the core, which we assume here collapses to form the initial seed BH. 

\begin{figure*}[t]
 \begin{minipage}[c]{0.5\textwidth}
\includegraphics[trim={0.8cm 0cm -0.8cm 0.0cm},clip,width=1.0\textwidth]{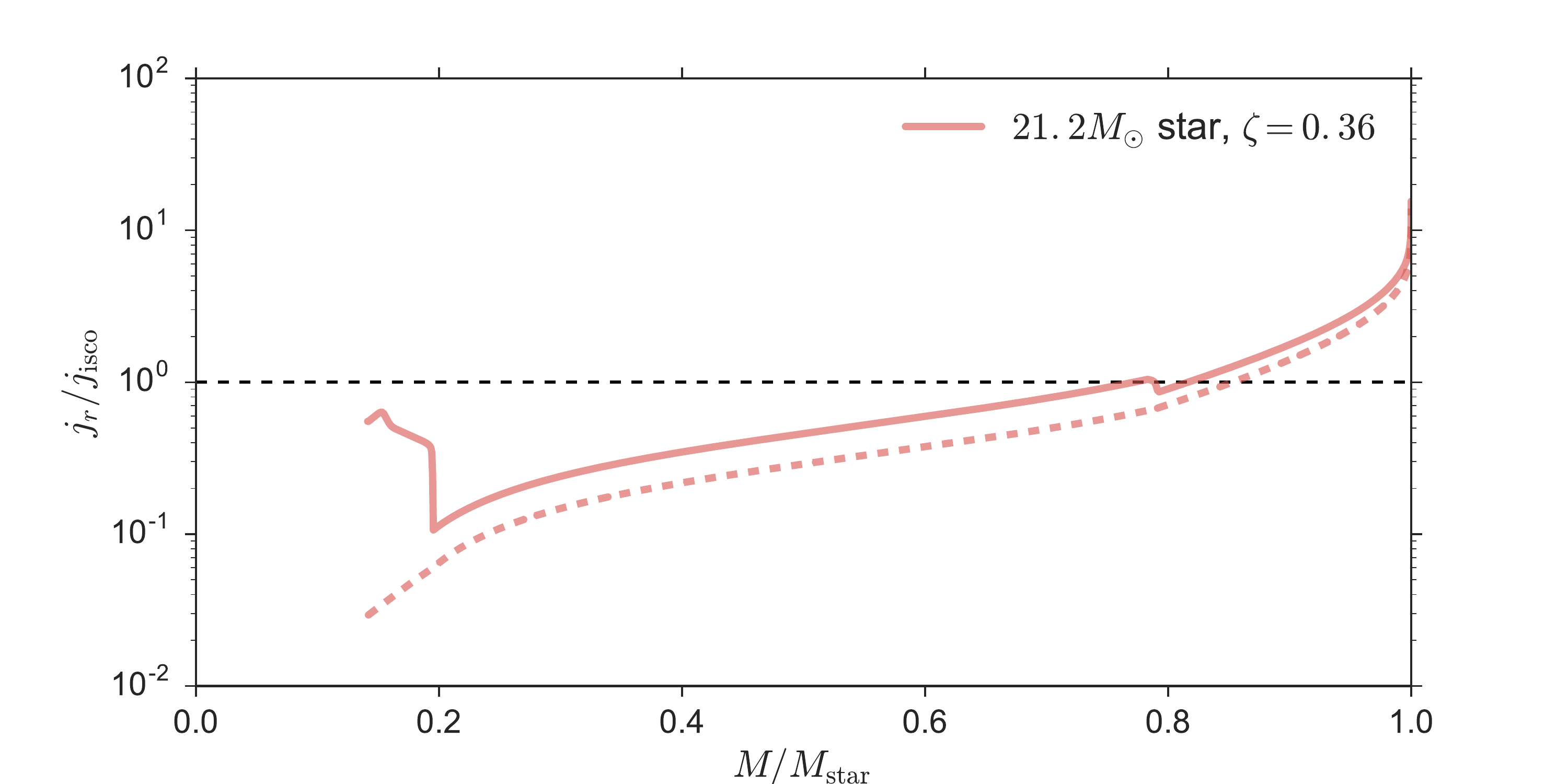}
\end{minipage}
 \begin{minipage}[c]{0.5\textwidth}
\includegraphics[trim={0.8cm 0cm -0.8cm 0.0cm},clip,width=1.0\textwidth]{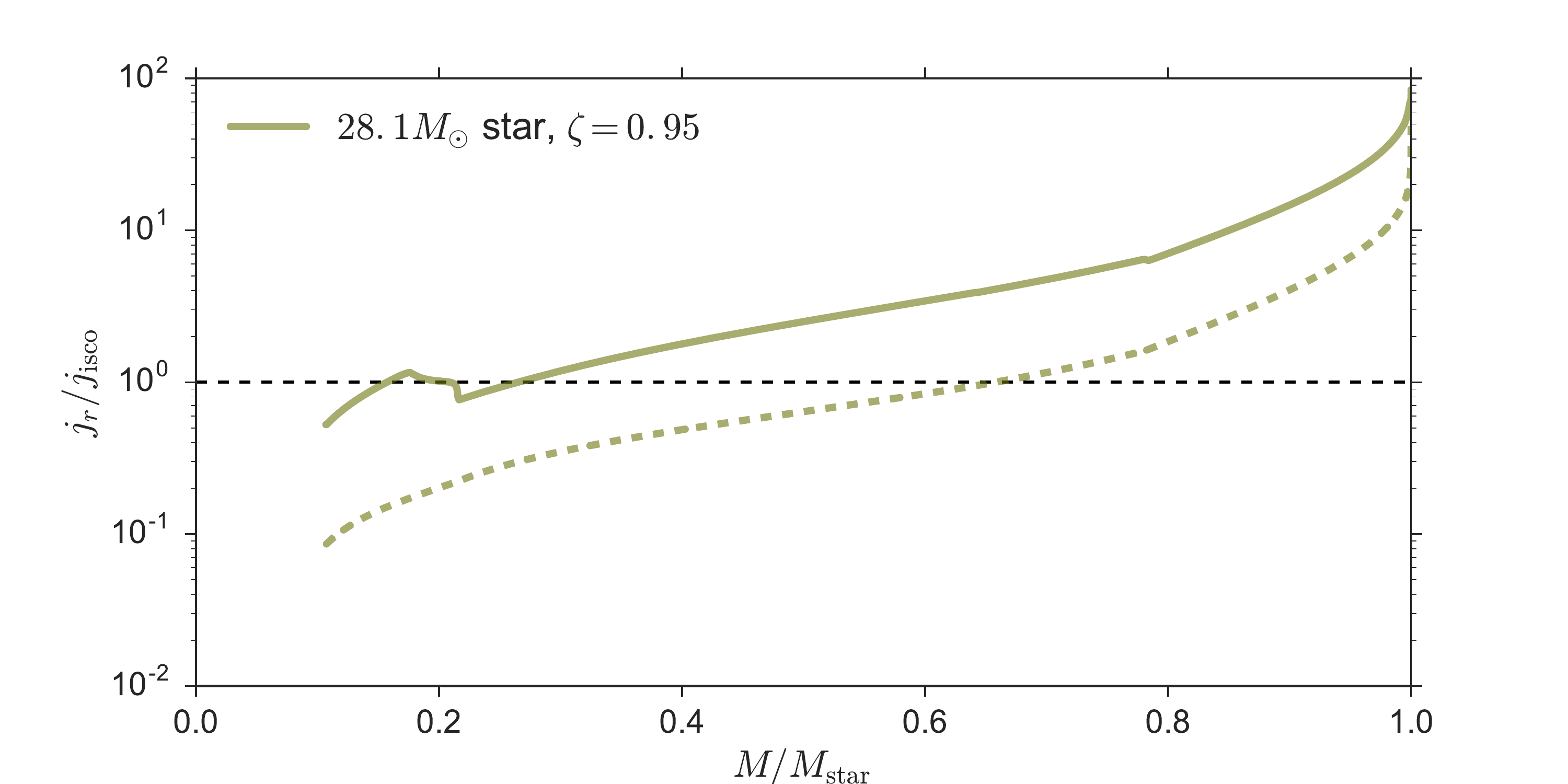}
\end{minipage}

 \begin{minipage}[c]{0.5\textwidth}
\includegraphics[trim={0.8cm 0.0cm -0.8cm 1.4cm},clip,width=1.0\textwidth]{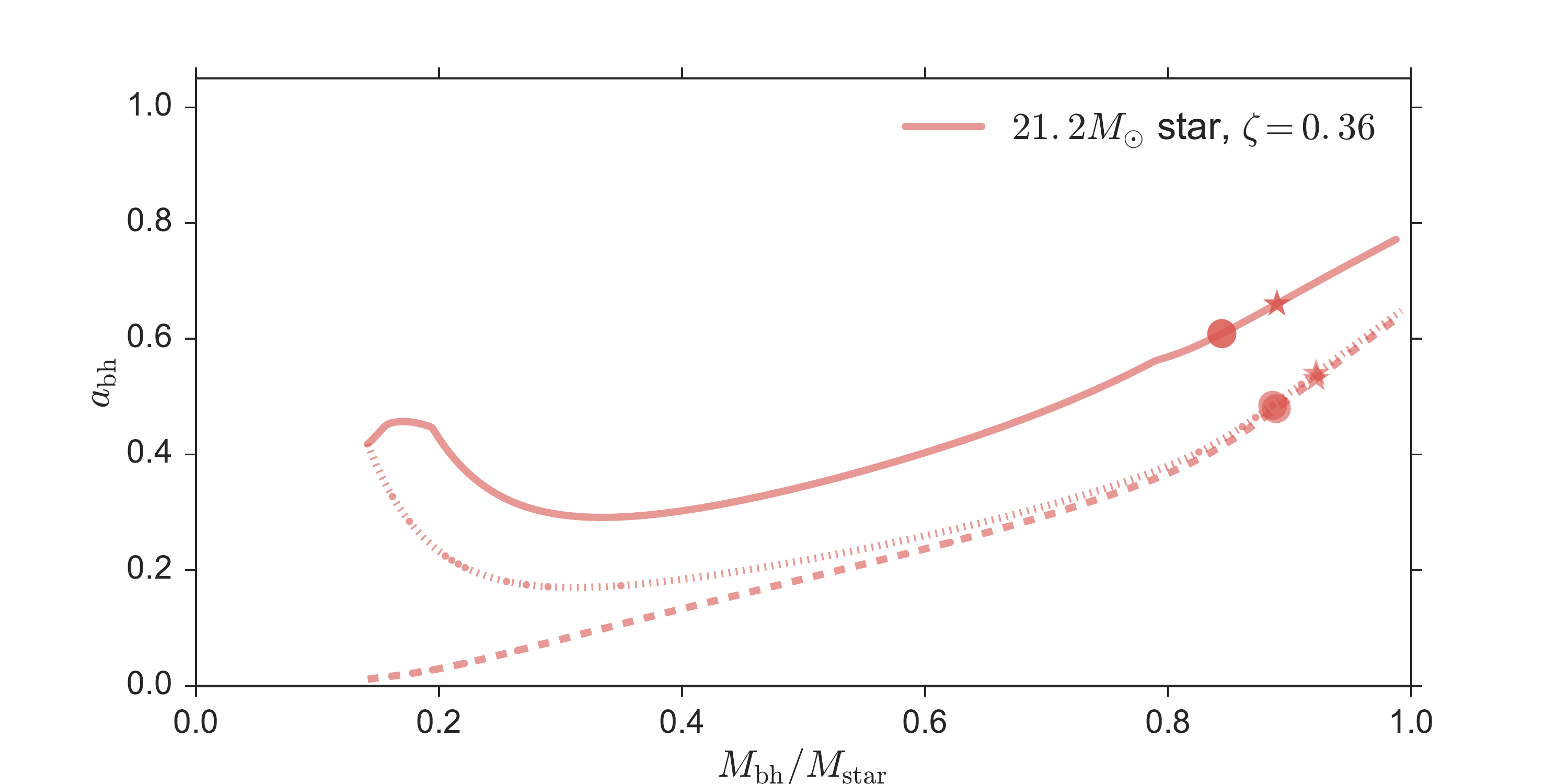}
\end{minipage}
 \begin{minipage}[c]{0.5\textwidth}
\includegraphics[trim={0.8cm 0.0cm -0.8cm 1.4cm},clip,width=1.0\textwidth]{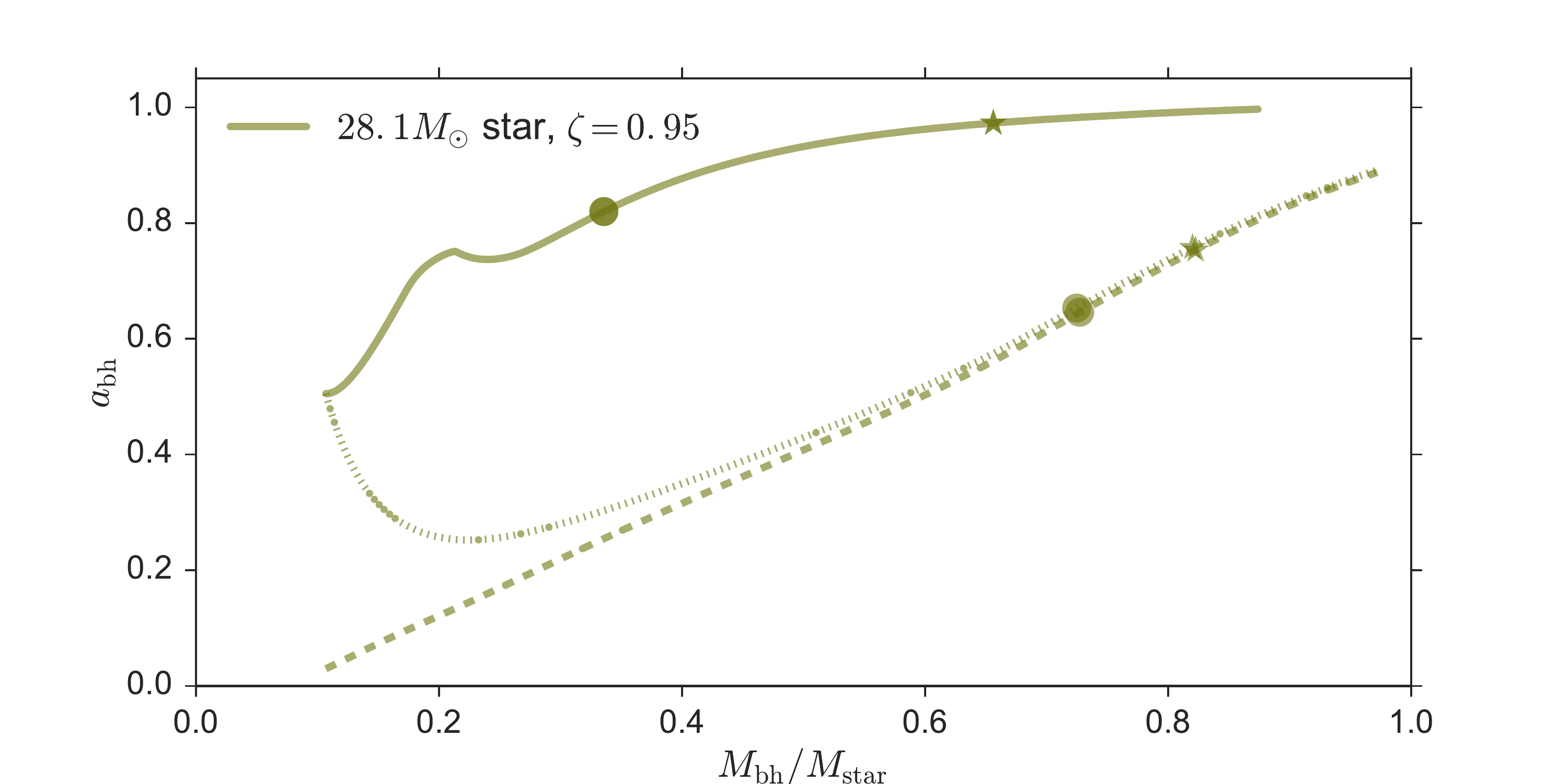}
\end{minipage}
\caption{Top panel: The angular momentum distributions for differential and rigid-body rotation ({\it solid} and {\it dashed} lines, respectively) for the $21M_{\odot}$ and $28M_{\odot}$ models ({\it left} and {\it right} panels, respectively). Bottom panel: The spin evolution of the BH as it grows  via accretion. In addition to the two models shown in the {\it top} panels, a third model is included ({\it dotted} line) in which we assume rigid body rotation outside the core and we preserve the rotation profile of the core. 
The circles and stars  correspond to the final  mass and spin of the BH for the feedback prescription without losses and with losses, respectively.}
 \vspace{0.4cm}
\label{fig:Diff_RB}
\end{figure*}

In the top panel of Figure \ref{fig:Diff_RB} we compare the models from WH06 with models that assume rigid-body rotation, which have been constructed to have the same surface angular velocity, $\Omega(R_{\rm star})$ ({\it solid} and {\it dashed} lines, respectively). The models from WH06 show a mild decrease of $\Omega(r)$ in the outer-layers, which causes the rigidly rotating  models to have a lower total angular momentum content. This leads to a smaller amount of stellar material with $j>j_{\rm isco}$, and ultimately gives rise to a BH with a lower total spin ({\it bottom} panel of Figure \ref{fig:Diff_RB}). Furthermore, the larger rotation rates in the cores of WH06 models give rise to initial BH seeds that are rotating faster.  In addition to the two rotational prescriptions shown in the {\it top} panels of Figure \ref{fig:Diff_RB}, a third model is included in the bottom panels ({\it dotted} line), in which we preserve the rotation profile of the core and assume rigid body rotation outside of it. 

The circles and stars on the {\it bottom} panel of Figure \ref{fig:Diff_RB} show the  final  mass and spin for the newly formed BHs calculated using our two feedback prescriptions. The two rigid rotation models, while having initial BHs with different  spins, end up with identical BHs. That is, the angular momentum content of the outer layers  is large enough to erase any memory of the initial spin of the BH. The rigidly rotating models constructed to match the surface angular velocity of the WH06 models produce BHs that are more massive and spin at lower rates.  \\

The angular momentum evolution of stars is highly uncertain and the models used here should be taken as illustrative. In the following subsection we investigate the impact that  the total angular momentum content of the star has in determining the properties of the newly formed BH. In what follows we assume, for simplicity, that stars  have rigid-body rotation.

\subsection{Rigid-Body Rotation Models}
To explore the effects of stellar spin, we assume rigid-body rotation and explore $\zeta$ values from 0 to 1.  The parameter $\zeta$ determines the transitional mass coordinate above which the collapsing material is able to form an accretion disk and subsequently generate mechanical feedback. To evaluate how the energy injection is shared with the collapsing envelope, we use the two feedback prescriptions described above and take $\epsilon=10^{-4}$, $10^{-3}$ and $10^{-2}$.

\begin{figure}[h!]  
\includegraphics[trim={0.8cm 0.0cm 0.0cm 0.0cm},clip,width=0.49\textwidth]{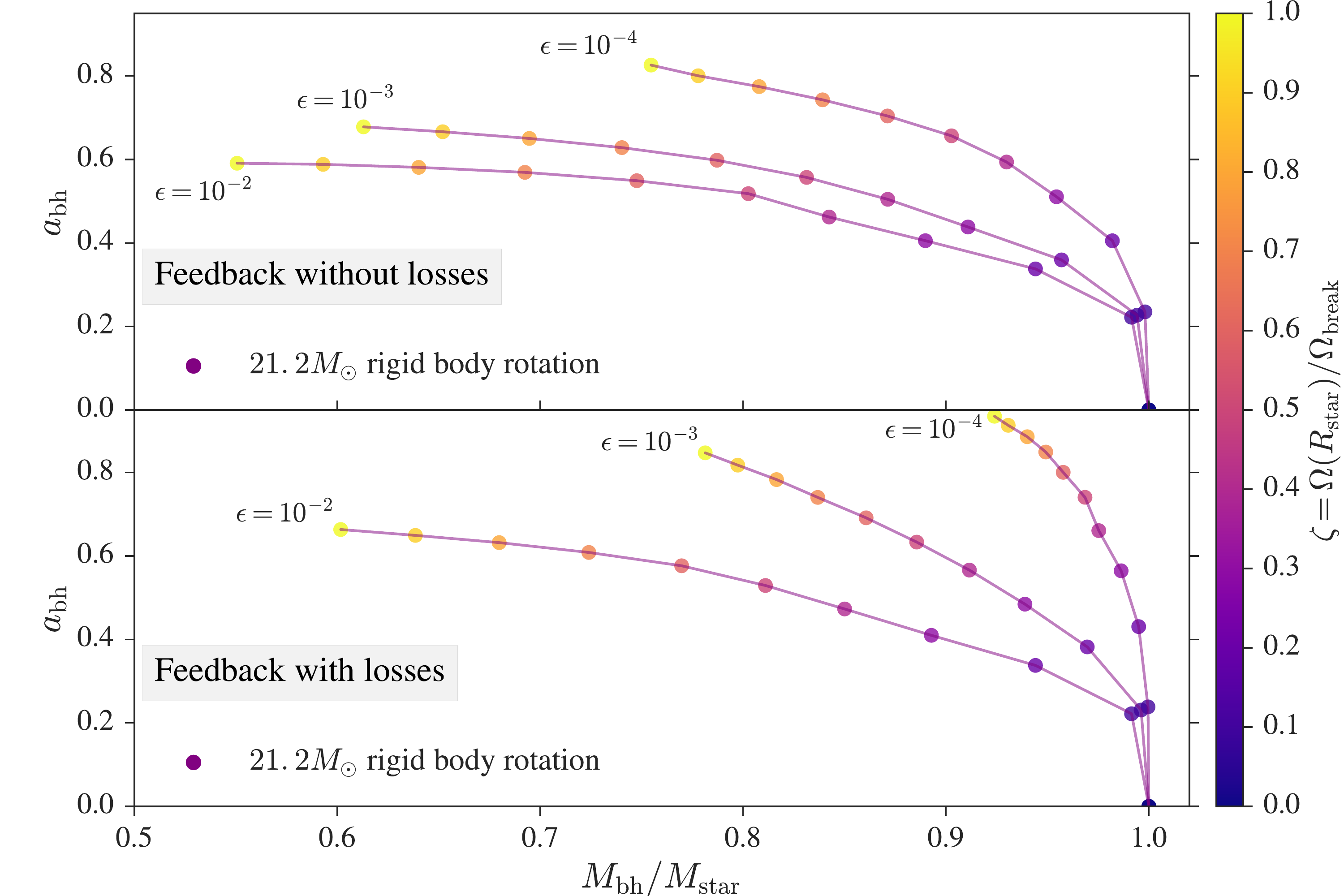}
\includegraphics[trim={0.8cm 0.0cm 0.0cm 0.0cm},clip,width=0.49\textwidth]{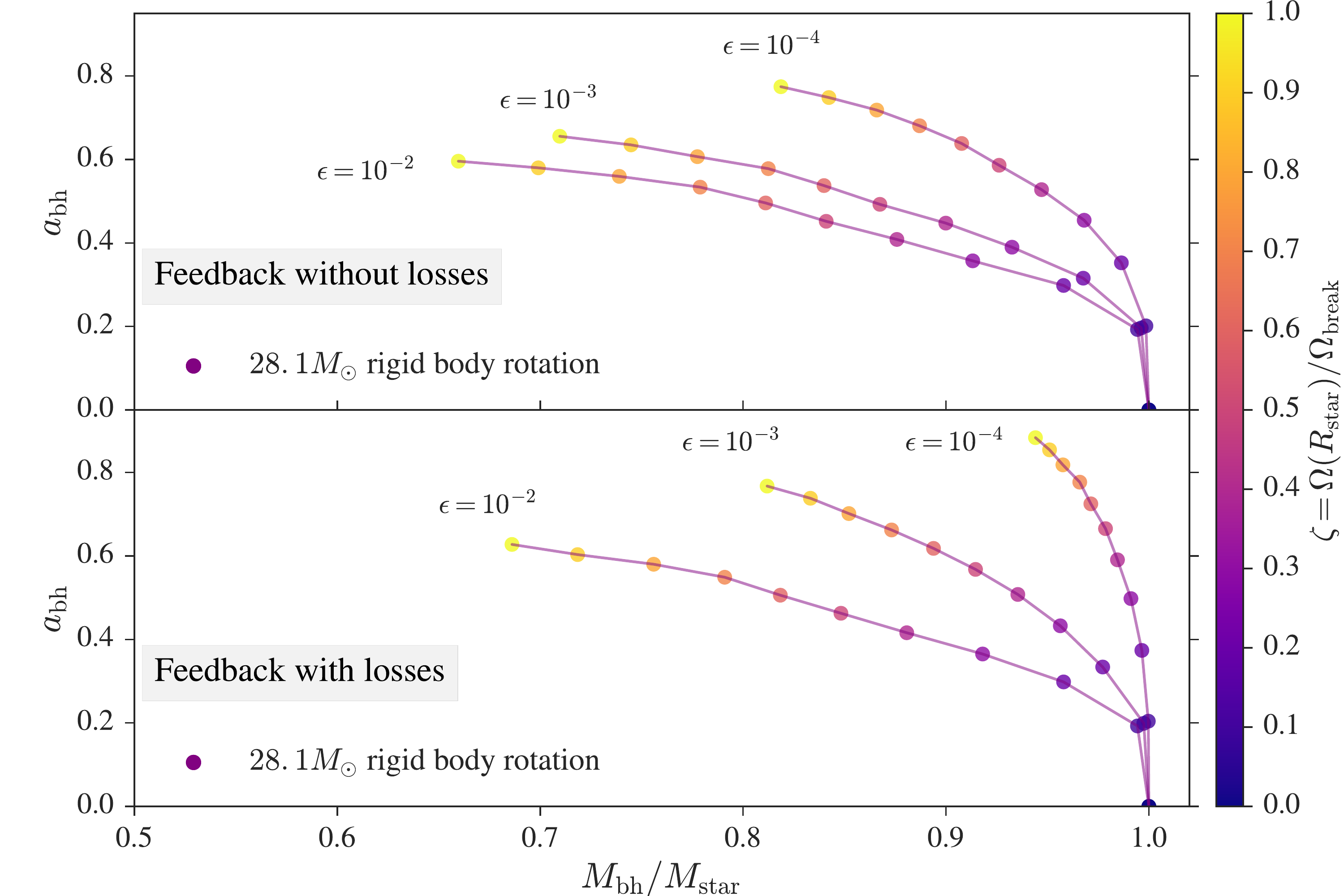}
\caption{The final spin and mass of the newly-formed BH obtained from the collapse of two different WH06 stellar models (21.2 and 28.1 $M_{\odot}$) with varying rotation rates. All models assume rigid-body rotation and explore different $\zeta$ values (color bar). We assume two feedback prescriptions and three different values of $\epsilon$. Different panels and our two feedback prescriptions. The formation of an accretion disk induces feedback, which in turn  imposes a limit to the fraction of the star that can collapse to form a BH.}
\label{fig:SpinOmega}
\end{figure}

The effects of varying stellar spins on the properties of the newly formed BH can be seen in Figure \ref{fig:SpinOmega} for two different stellar models: 21.2 and 28.1 $M_{\odot}$. Each panel depicts the dependence of $\zeta$ (color bar) and $\epsilon$ (three different curves) on the final spin and mass of the BH. In all our calculations, whenever an accretion disk is formed, feedback imposes a limit to the fraction of the star that can collapse to form a BH which in turn limits the final spin and mass that a BH can have. 

For low stellar spins (bottom right in all panels in Figure \ref{fig:SpinOmega}), the star collapses completely without forming an accretion disk. As a result, there is no feedback and the BH accretes the entire low angular momentum stellar material. As the angular momentum content of the star increases, the amount of material that is able to form an accretion disk augments and feedback becomes progressively more relevant at earlier times during the collapse. Under the circumstances, the final mass of the resulting BH decreases while the final spin increases (top left in all panels in Figure \ref{fig:SpinOmega}). As the feedback efficiency, $\epsilon$, decreases it takes longer to build up enough mechanical energy to unbind the envelope and, subsequently, more mass is allowed to fall into the BH.

\subsection{Accretion Disk Assembly and Evolution}
\begin{figure}

\includegraphics[trim={0.55cm 0.0cm 0.8cm 1.7cm},clip,width=0.49\textwidth]{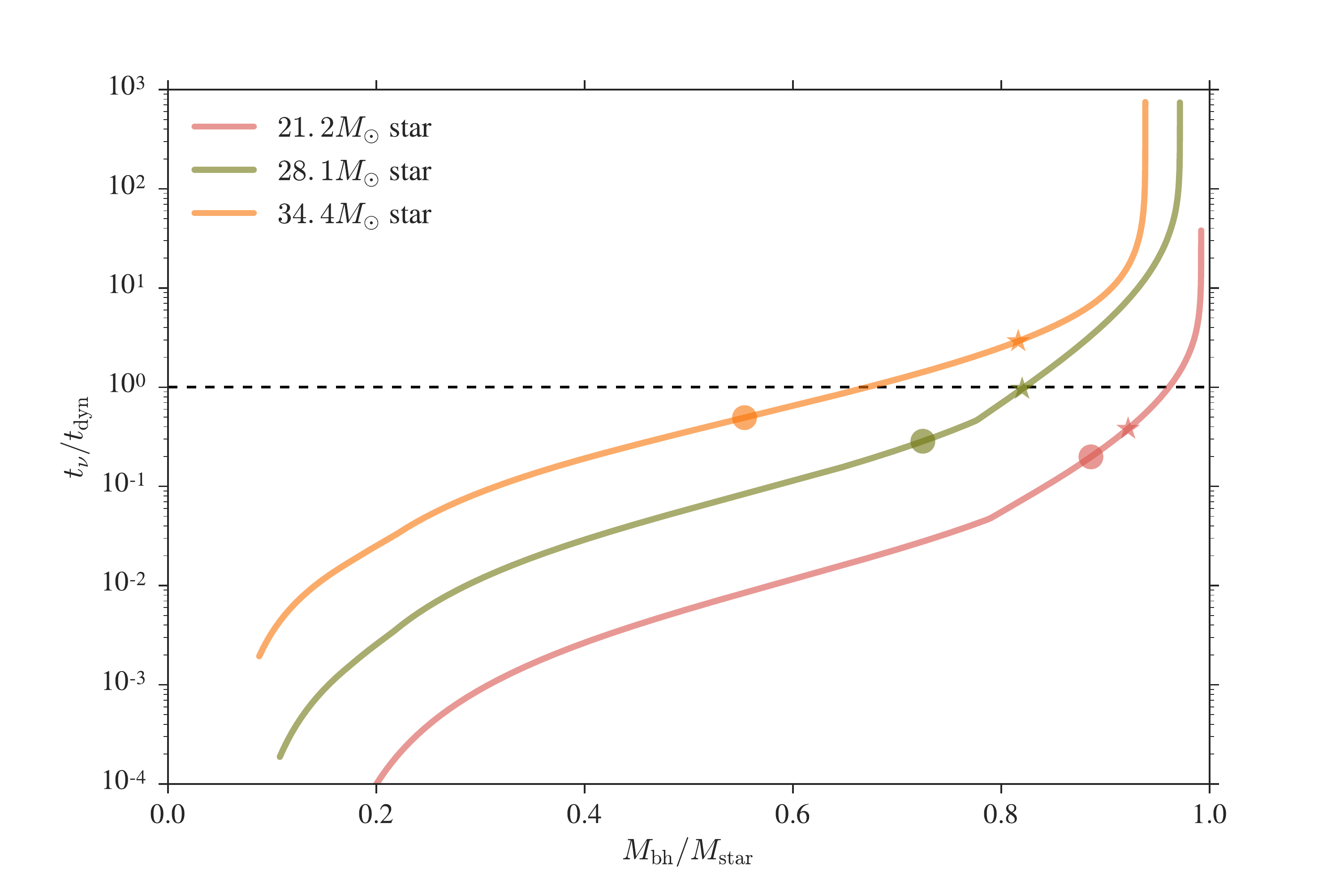}
\caption{The viscous and dynamical time scales ($t_{\nu}$ and $t_{\rm dyn}$, respectively) for the rigidly rotating models shown in Figure  \ref{fig:Diff_RB} assuming $\alpha (H/R)^2=10^{-3}$. The circles and stars indicate the mass of the BH at which the star gets unbound by the two differemt feedback prescriptions used here, which are referred to as without losses and with losses, respectively.}
 
\label{fig:Viscoustime}
\end{figure}

The calculations shown in Figure \ref{fig:SpinOmega} have been constructed under the assumption that the viscous delay time from accretion formed from the collapse of a shell is shorter than the time it takes the next shell to collapse. Figure \ref{fig:Viscoustime} shows the evolution of the ratio between the accretion disk's viscous timescale and the shell's dynamical time scale for the rigidly rotating models shown in Figure  \ref{fig:Diff_RB}. These models have been constructed to have the same surface angular velocity as the WH06 models.
The entire 21.2 and 28.1 $M_\odot$  stellar models get unbound before $t_{\nu}/t_{\rm dyn}>1$ for  both feedback prescriptions. In this case,  there is no significant amount of mass accumulation taking place within the disk and the mass and spin of the resultant BH are consistently calculated until the collapse of the star is halted. However, this is not entirely the case for the more massive model. Here, the disk will accumulate a small fraction of mass before the envelope is unbound. The exact result of this is not trivial, given the close interplay between the feedback and the resultant accretion.  The overall result will be a slightly different evolution of the spin and mass than the one calculated here using this simple model. Having said this, the simple model can be used as a guide to understand the role of feedback in the production of BH from direct collapse.

\section{Discussion}
\subsection{Feedback and Implications for BH Formation}
As discussed  here, our calculations show that accretion feedback can play a key role in shaping the final properties of BHs formed from the direct collapse of rotating pre-SN progenitors. Stars with rotation rates that are large enough to form an accretion disk are expected to unbind their outer layers and produce BHs with only a fraction of the total mass and angular momentum of the progenitor star. 

\begin{figure}[!h]
 \includegraphics[trim={0.55cm 0.0cm 0.8cm 0.0cm},clip,width=0.49\textwidth]{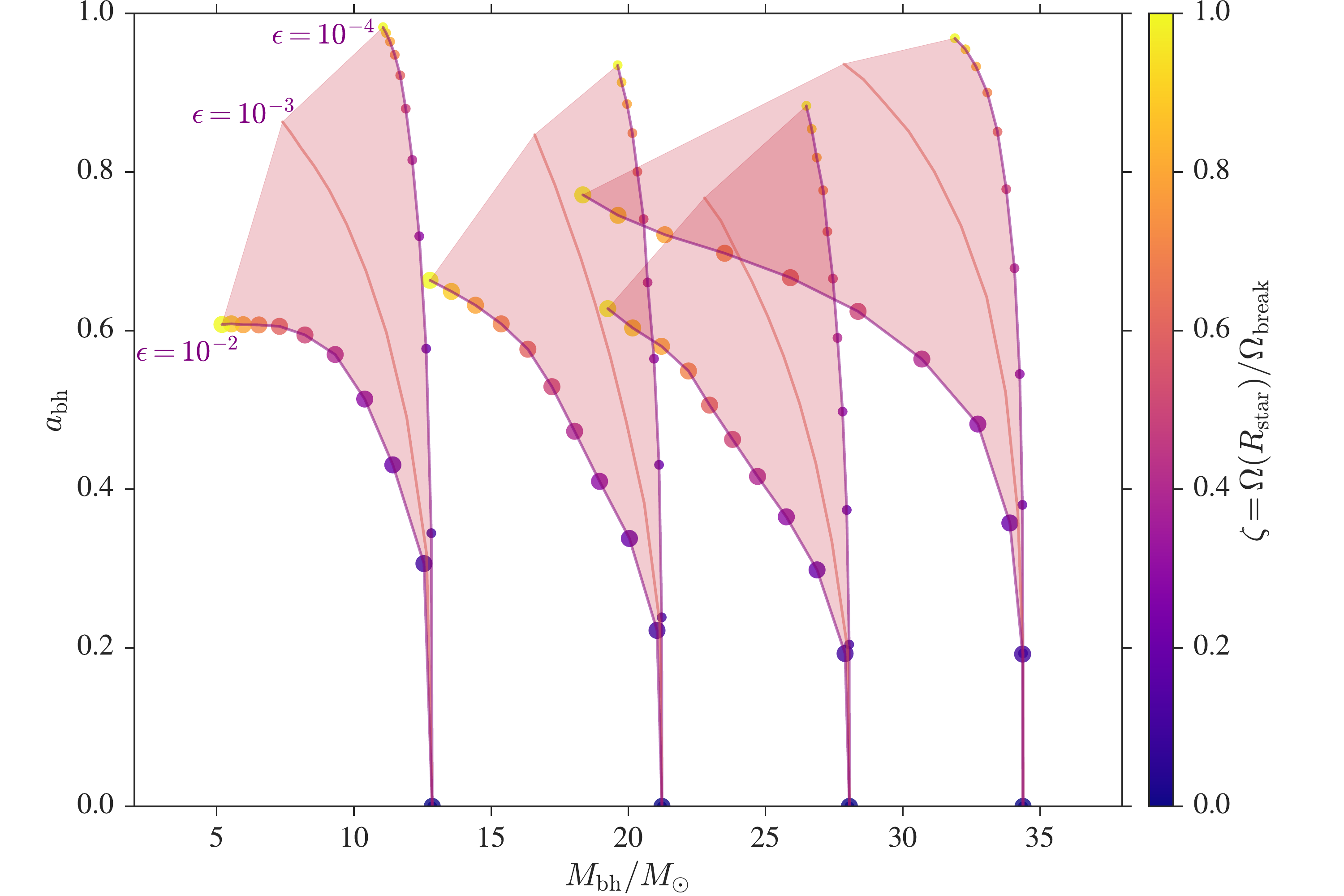}
\caption{The final spins and masses of  newly-formed BHs obtained from the collapse of four different WH06 stellar model profiles with varying rotation rates. Similar to Figure \ref{fig:SpinOmega}, shown are the masses and spins of the resulting BHs as function of $\zeta$ and $\epsilon$. We have assumed here the feedback prescription with losses.}

\label{fig:FeedbackWH06}
\end{figure}

Assuming rigid-body rotation, the expected final mass and spin of a newly formed BH produced by the collapse of pre-SN progenitor stars can be easily calculated as a function of $\zeta$. Figure \ref{fig:FeedbackWH06} shows the mass and spin of the resulting BH as a function of the stellar rotation rate and $\epsilon$. For slowly rotating stars, the entire mass is accreted by the BH. For inefficient feedback, a star of mass $M_{\rm star}$ should follow an almost vertical line in the $a_{\rm bh}-M_{\rm bh}$ plane as $\zeta$ increases. However, a mild feedback efficiency will yield a sizable reduction of the final mass of the BH. The individual red shaded regions in Figure \ref{fig:FeedbackWH06} correspond to particular stellar models and indicate the resulting range of spins and masses of the resulting BHs within our selected ranges of $\zeta$ and $\epsilon$. The results from differentially  rotating progenitors are expected to be qualitatively similar, although rigid-body rotation models will generally produce more massive BHs with lower spins than the ones derived from the WH06 progenitor models (Figure~\ref{fig:Diff_RB}).  

Efficient feedback  will invariably reduce the expected  masses and spins of BHs formed from direct collapse. This has two important consequences. First, it implies that there is a critical spin for which a BH is able to acquire the entire mass and angular momentum of the progenitor star without forming an accretion disk. This critical spin depends on the star's angular momentum distribution which, for stars with rigid-body rotation, is determined by the moment of inertia and the angular velocity. For the stellar progenitors used in this work, the critical spin obtained without forming an accretion disk ranges from 0.1 to 0.14, which corresponds to rotation rates $\zeta \approx 0.07$. Second, it limits the formation of very massive and rapidly rotating BHs (Figures \ref{fig:SpinOmega} and \ref{fig:FeedbackWH06}).

The role of accretion feedback could have profound consequences for the formation of rapidly rotating BHs through direct collapse in high mass X-ray binaries (HMXRB), further accentuating the need of alternative spin up mechanisms \citep{Batta_2017,Qin_Fragos_2019,Sophie2018}. However, such systems can also be used to shed some light on the efficiency of the feedback mechanism during BH formation. Specially if one considers the recent work by \cite{Qin_Fragos_2019}, which is able to produce close binaries that reproduce the properties of the observed HMXRB, with rapidly rotating He stars with masses between $\approx15M_{\odot}$ and $\approx20M_{\odot}$. In order for such He stars to produce rapidly spinning BHs with masses observed in HMXRBs (between $\approx10M_{\odot}$ and $\approx 16M_{\odot}$), the accretion feedback efficiency must be relatively low to avoid a significant loss of mass and spin as shown in Figure \ref{fig:SpinMass_epsilon}.

\subsection{Stellar Rotation  and  LIGO BHBs}
LIGO observations of the mass-weighted angular momentum $\chi_{\rm eff}$
perpendicular to the orbital plane  have been argued to provide stringent constraints on progenitor channels. Given our limited knowledge about stellar rotation of massive stars, it is difficult to accurately constraint the spin of BHs formed in isolation. The models shown in Figure \ref{fig:FeedbackWH06} can be compared to  observations of LIGO BHBs provided that meaningful constraints can be placed on the angular momentum content of the stellar progenitors. 

While BHBs assembled dynamically in dense stellar systems
are expected to form in isolation, BHBs formed from
the classical binary field formation channel are predicted to
yield spins with correlated orientations \citep{Tutukov_1993,Belczynski_2016}. In this progenitor
route, a wide massive binary undergoes a number of mass
transfer episodes leading to a tight binary comprised of a helium
star and a BH.

As discussed in \citet{Qin_Fragos_2018}, the spin from the first-born BH is expected to be negligible. The angular momentum content of the secondary BH, on the other hand, is closely related to that of the helium pre-SN progenitor, which can be effectively driven to synchronization by tidal torques during its He burning phase \citep{Qin_Fragos_2018,Zaldarriaga_2018} for a limited range of binary separations $d_{\rm RL}<d<d_{\tau}$. The lower limit $d_{\rm RL}$ is determined by the binary's minimum separation before Roche lobe overflow (RLOF) \citep{Eggleton_1983}, and the upper limit $d_{\tau}$ by the maximum separation within which synchronization is achieved over the helium star lifetime. In this scenario, the built-in angular velocity of the stellar progenitor, $\zeta$, can thus be determined by the orbital separation, which increases as the binary separation decreases. 

However, tidal synchronization is no longer efficient after the He burning phase. This is because the star's evolution time scale is   too short to allow tidal synchronization to occur. Thus, the collapse of the core on every ensuing burning stage will reduce its moment of inertia and increase its angular velocity accordingly,  ultimately changing the star's angular momentum distribution, which will remain mostly unaffected by the binary companion. These has two possible outcomes. First, if angular momentum transport processes within the star are efficient, rigid body rotation (constant angular velocity) will be eventually restored, and the rotation rate $\zeta$ will be higher than the one during the synchronized He burning stage. Second, if angular momentum transport within the star is inefficient,  differential rotation will endure, and the star's core will be rotating at a significantly higher rate than the outer layers (as observed in the WH06 models). Therefore, the star's rotation rate $\zeta$ right before collapse will not be the same as the one determined by synchronization; it will be higher throughout the entire star (for rigid body rotation) or higher in the core as in the WH06 models. Irrespective of the transport efficiency, as long as mass loss is not important, the angular momentum content of the star should remain unchanged, which for slow rotation rates ($\zeta\lesssim0.1$) will yield a BH with a similar spin as that of the stellar progenitor. 

\begin{figure}[!h]
\includegraphics[trim={0.55cm 0.0cm 0.8cm 0.0cm},clip,width=0.49\textwidth]{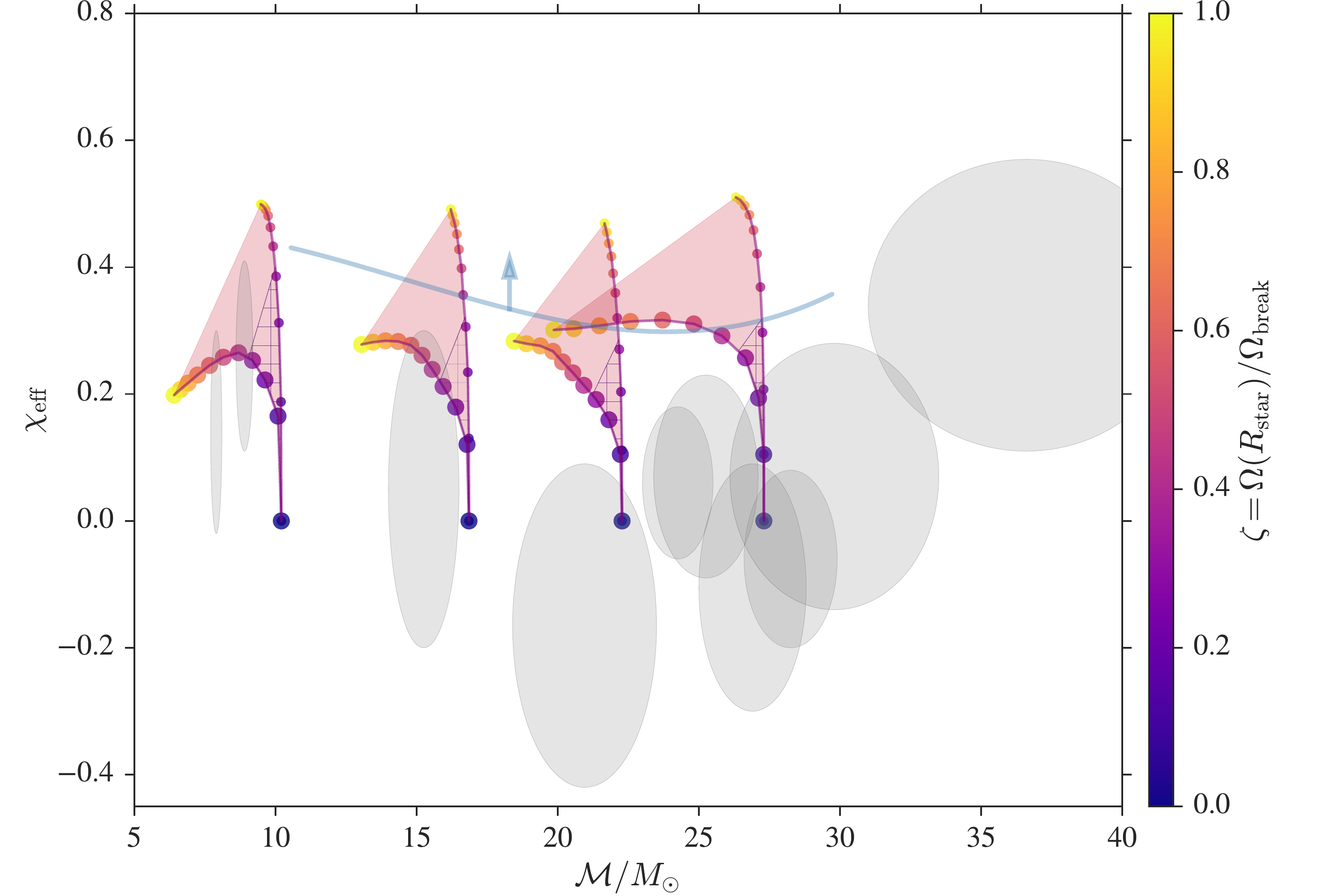}
\caption{The mass-weighted angular momentum $\chi_{\rm eff}$ and chirp mass $\mathcal{M}$ for BHBs formed in the classical binary field progenitor channel. The properties of the second-born BH are calculated using the results shown in   Figure \ref{fig:FeedbackWH06}. The spin of the first-born BH is set to zero while it mass is given by $M_{\rm bh}q^{-1}$, where $q=1.2$ as inferred in GW150914 and GW170814. The gray shaded regions show the 90\% credibility intervals of LIGO events. The purple hatched regions illustrate  the expected range $\zeta$ for tidally synchronized binaries, limiting the effective spin to $\chi_{\rm eff}<0.4$. The blue curve shows the expected  $\chi_{\rm eff}$ and $\mathcal{M}$ for BHBs formed from the chemically homogeneous evolution channel. This curve has been calculated assuming two equal mass WH06 progenitor binaries with $\zeta>0.2$.}
\label{fig:FeedbackLIGO}
\end{figure}

In Figure \ref{fig:FeedbackLIGO} we show the results for BHBs arising from the classical binary field formation channel discussed above. To calculate the  effective spin parameter $\chi_{\rm eff}$ and chirp mass $\mathcal{M}$ of a corresponding BHB, we postulate that the spin and mass of the second-born BH are accurately described by the results shown in Figure~\ref{fig:SpinOmega} and assume that the first-born BH is not spinning and its mass is given by $M_{\rm bh}q^{-1}$. Here $q$ is the mass ratio, which is taken to be $q=1.2$ (similar to the mass ratio inferred in GW150914 and GW170814). The gray shaded regions in Figure \ref{fig:FeedbackLIGO} correspond to the 90\% credibility intervals of LIGO BHBs \citep{LIGO_Catalog2018}. 

Table \ref{tab:tidallylocked_BHB} shows the range of critical separations at which our BH-helium star binaries can be driven to synchronization by tidal torques \citep{Zaldarriaga_2018}, along with the corresponding rotation rates $\zeta$. In order to ensure synchronization, the orbital separation must be $d\leq d_{\tau}$, yielding $\zeta(d_{\tau})\sim 10^{-2}$. In order to avoid RLOF, the minimum binary separation needs to be $d_{\rm RL}=2.53 R_{\rm star}$ for all stars, which translates into $\zeta(d_{\rm RL})=0.336$. The purple hatched regions in Figure \ref{fig:FeedbackLIGO} show the expected $\chi_{\rm eff}$ and $\mathcal{M}$ for such tidally synchronized binaries, which should be considered as approximate values. The obtained effective spins are nonetheless  consistent with observations ($\chi_{\rm eff}\lesssim0.4$), and are not expected to vary significantly  since the angular momentum content of the star is likely to remain unchanged  after tidal synchronization.

\begin{table}[!h]
    \centering
    \begin{tabular}{c|c|c|c|c}
     $M_{\rm star}/M_{\odot}$ & $d_{\tau}/R_{\rm star}$  & $d_{\rm RL}/R_{\rm star}$ & $\zeta(d_{\tau})$ & $\zeta(d_{\rm RL})$ \\\hline
          12.9 & 12.50 & 2.53 & 0.030 & 0.336\\
          21.2 & 21.45 & 2.53 & 0.014 & 0.336\\
          28.1 & 19.55 & 2.53 & 0.016 & 0.336\\
          34.4 & 35.66 & 2.53 & 0.006 & 0.336\\

    \end{tabular}
    \caption{Properties of tidally synchronized BH-helium star binaries shown in Figure~\ref{fig:FeedbackLIGO}. Shown (from left to right) is the mass of the He star, the maximum binary separation allowed for synchronization, the minimum separation to avoid RLOF, and the corresponding rotation rate $\zeta$ for such separations.}
    \label{tab:tidallylocked_BHB}
\end{table}

As discussed in \citet{Zaldarriaga_2018}, the origin of LIGO BHBs through the classical field formation channel requires one or both of the BHs to have intrinsically low spins, or be anti-aligned. The inclusion of feedback, as we have demonstrated here, will help in producing slowly spinning BHs, even in scenarios where the second BH comes from the collapse of a rapidly spinning helium star \citep{Zaldarriaga_2018}. Also plotted in Figure \ref{fig:FeedbackLIGO} is our estimate of $\chi_{\rm eff}$ and $\mathcal{M}$ for BHBs arising from the chemically homogeneous evolution channel (blue curve). This constraint has been calculated using two equal mass WH06 progenitors with $\zeta>0.2$, as required for effective mixing to occur \citep{Yoon2006} and assuming $\epsilon=10^{-2}$.

In closing, our work shows that accretion feedback, regardless of the formation channel, will invariably reduce the expected masses and spins of BHs formed from direct collapse. Moreover, it prevents the formation of  massive, rapidly spinning BHs, since strong feedback is expected to be crucial for rapidly rotating progenitors. Thus, the most massive LIGO BHs must come from slowly rotating stars, or from heavier more rapidly spinning progenitors.  Obviously, the calculations highlighted  here are only sketchy and should be taken as an order of magnitude  at present. Much of our thinking regarding the nature of black hole feedback in direct stellar collapse is conjecture and revolves primarily  around different prejudices as to how magnetic, three
dimensional flows behave in a strong gravitational field. What is more valuable, though considerably harder to achieve is to refine models like the ones presented here to the point of making quantitative
predictions, and to evaluate and interpret LIGO observations so as to constrain and refute these theories.
What we can hope of the formalism presented in this paper  is that it will assist us in this venture.

\section*{Acknowledgements}
We thank I. Mandel, J. Murphy, J. McKinney, W. Farr, T. Fragos, J. Andrews, and S. Schr\o der for useful conversations. We credit the Packard Foundation and the DNRF for support. We are grateful to the Kavli Foundation and the DNRF for sponsoring the 2017 Kavli Summer Program, where part of this work was carried out.

\acknowledgments

\bibliography{BHBinaries}

\end{document}